\documentclass[aps,prd,twocolumn,showpacs,superscriptaddress,groupedaddress,floatfix]{revtex4}  % for review and submission

\usepackage{graphicx}  % needed for figures
\graphicspath{ {./plots/} }
\usepackage{bm}        % for math
\usepackage{amssymb}   % for math
\usepackage[colorlinks=true]{hyperref}

\usepackage{pifont}% http://ctan.org/pkg/pifont
\newcommand{\cmark}{\textcolor{green}{\ding{51}}}%
\newcommand{\xmark}{\textcolor{red}{\ding{55}}}%
\newcommand{\qmark}{\textcolor{blue}{\textbf{?}}}

\newcommand{\floor}[1]{\lfloor #1 \rfloor}
\newcommand{\comment}[1]{}

\newcommand{\lr}[1]{ \left( #1 \right) }
\newcommand{\lrs}[1]{ \left[ #1 \right] }
\newcommand{\lrc}[1]{ \left\{ #1 \right\} }
\newcommand{\vev}[1]{ \langle \, #1 \, \rangle }

\newcommand{\Tr}{ {\rm Tr} \, }
\newcommand{\tr}{ {\rm Tr} \, }

\newcommand{\ket}[1]{ \, | #1 \rangle }
\newcommand{\bra}[1]{ \langle #1 | \, }

\renewcommand{\det}[1]{ {\rm det} \left( #1 \right) }

\newcommand{\sign}{ {\rm sign} \,  }

\newcommand{\expa}[1]{ \exp{\left( #1 \right)} }

\begin{document}

\title{Numerical study of chiral plasma instability within the classical statistical field theory approach}

\author{P. V. Buividovich}
\email{Pavel.Buividovich@physik.uni-regensburg.de}
\affiliation{Institute of Theoretical Physics, University of Regensburg,
D-93053 Germany, Regensburg, Universit\"{a}tsstrasse 31}

\author{M. V. Ulybyshev}
\email{Maksim.Ulybyshev@physik.uni-regensburg.de}
\affiliation{Institute of Theoretical Physics, University of Regensburg,
D-93053 Germany, Regensburg, Universit\"{a}tsstrasse 31}
\affiliation{Institute for Theoretical Problems of Microphysics, Moscow State University, Moscow, 119899 Russia}

\date{May 23rd, 2016}

\begin{abstract}
 We report on a numerical study of real-time dynamics of electromagnetically interacting chirally imbalanced lattice Dirac fermions within the classical statistical field theory approach. Namely, we perform exact simulations of the real-time quantum evolution of fermionic fields coupled to classical electromagnetic fields, which are in turn coupled to the vacuum expectation value of the fermionic electric current. We use Wilson-Dirac Hamiltonian for fermions, and non-compact action for the gauge field. In general, we observe that the backreaction of fermions on the electromagnetic field prevents the system from acquiring chirality imbalance. In the case of chirality pumping in parallel electric and magnetic fields, electric field is screened by the produced on-shell fermions and the accumulation of chirality is hence stopped. In the case of evolution with initially present chirality imbalance, axial charge tends to transform to helicity of electromagnetic field. By performing simulations on large lattices we show that in most cases this decay process is accompanied by the inverse cascade phenomenon which transfers energy from short-wavelength to long-wavelength electromagnetic fields. In some simulations, however, we observe a very clear signature of inverse cascade for the helical magnetic fields which is not accompanied by the axial charge decay. This suggests that the relation between inverse cascade and axial charge decay is not as straightforward as predicted by the simplest form of anomalous Maxwell equations.
\end{abstract}

\maketitle

\section{Introduction and brief summary}
\label{sec:intro}

 Over the past few decades, real-time instability of the system of chiral fermions coupled to dynamical gauge fields has been attracting a lot of attention in various fields of physics, ranging from astrophysics to condensed matter physics. This instability manifests itself in the decay of the initial imbalance between the densities of the left- and right-handed fermions at the expense of the generation of magnetic fields with nonzero magnetic helicity (or, in other words, winding number of magnetic flux lines).  In the astrophysical context, the phenomenon of chiral plasma instability is actively discussed as the mechanism responsible for the generation and enhancement of primordial magnetic fields \cite{Shaposhnikov:97:1,Semikoz:11:1,Boyarsky:12:1,Tashiro:12:1,Giovannini:13:1} as well as for the transfer of magnetic field energy from short to cosmological scales \cite{Boyarsky:12:1,Boyarsky:15:1,Sigl:15:1}.

 In the context of condensed matter physics, chiral plasma instability was initially discussed and experimentally detected as the so-called helical instability of liquid ${}^3$He \cite{VolovikHeliumDroplet}. It was also considered as a mechanism of spontaneous magnetization of topological magnetic insulators \cite{Ooguri:12:1}. In experiments in which chirally imbalanced Weyl semimetal states are created from Dirac semimetals by applying parallel electric and magnetic fields \cite{Kharzeev:14:1,Kim:13:1,Xiong:15:1} chiral plasma instability might manifest itself in the spontaneous emission of circularly polarized terahertz-range electromagnetic radiation \cite{Kharzeev:15:1}.

 In heavy-ion collisions, chiral plasma instability might lead to enhanced emission of circularly polarized soft photons \cite{Kharzeev:15:1}. It should be also important for the correct estimate of the lifetimes of chirality imbalance and magnetic fields \cite{Yamamoto:13:1,Manuel:15:1}. However, the estimates of \cite{Tuchin:14:1} suggest that in heavy-ion collisions the volume and the lifetime of the quark-gluon plasma might be too small for the instability to develop.

 The origin of this instability of chirally imbalanced Dirac fermions is the Chiral Magnetic Effect (CME) \cite{Kharzeev:08:2,Kharzeev:09:1} - electric current flowing parallel to the magnetic field in the presence of chirality imbalance. Within the linear response approximation the contribution of CME to the electric current is
\begin{eqnarray}
\label{cme_current}
 \vec{j}_{CME} = \sigma_{CME} \, \vec{B} .
\end{eqnarray}
The commonly quoted value for the chiral magnetic conductivity $\sigma_{CME}$ is $\sigma_{CME} = \frac{\mu_A}{2 \pi^2}$, where $\mu_A$ is the so-called chiral chemical potential which parameterizes the difference between the Fermi levels of right- and left-handed fermions and hence also the total axial charge of the system. The value of $\sigma_{CME}$, however, strongly depends on frequency $w$ and wave vector $\vec{k}$ of electromagnetic field, and, in the limit of constant and homogeneous magnetic field, on the way in which the limits $w \rightarrow 0$ and $k \rightarrow 0$ are taken \cite{Kharzeev:09:1,Ren:11:1,Yee:14:1,Buividovich:13:8,Landsteiner:11:2,Landsteiner:12:1,Buividovich:15:1}.

 In order to see how the CME current (\ref{cme_current}) can lead to instability, one can insert it into the classical Maxwell equations, along with the conventional Ohmic current $\vec{j} = \sigma \vec{E}$, where $\sigma$ is the electric conductivity. Assuming the unbroken translational invariance both in time and space, we can write these so-called anomalous Maxwell equations \cite{Shaposhnikov:97:1,Boyarsky:12:1,Yamamoto:13:1,Tashiro:12:1,TorresRincon:13:1,TorresRincon:14:1,Tuchin:14:1,Kharzeev:15:1,Shovkovy:16:1} in frequency-momentum space as
\begin{eqnarray}
\label{anomalous_maxwell}
 i w \vec{B} = - i \vec{k} \times \vec{E}, \quad
 i w \vec{E} =   i \vec{k} \times \vec{B} - \sigma \, \vec{E} - \sigma_{CME} \, \vec{B} .
\end{eqnarray}
From these equations we find the following four-branch dispersion relation for transversely polarized plane waves with the wave vector $\vec{k} = k \, \vec{e}_3$ \cite{Tuchin:14:1}:
\begin{eqnarray}
\label{anomalous_maxwell_dispersion}
 w_{s,r} = \frac{i \sigma}{2} + s \sqrt{k^2 + r \sigma_{CME} k - \frac{\sigma^2}{4} } ,
\end{eqnarray}
where $s = \pm 1$ and $r = \pm 1$ label different branches of the dispersion relation. The corresponding polarization vectors $\epsilon_r = 2^{-1/2} \, \lr{1, -i r, 0}$ for the electric field $\vec{E}$ correspond to circularly polarized waves with opposite helicities (handedness) for opposite $r$.

 While for nonzero electric conductivity $\sigma$ the imaginary part of $w$ in (\ref{anomalous_maxwell_dispersion}) is always positive and hence corresponds to decaying plane waves, nonzero chiral magnetic conductivity can also lead to exponentially growing solutions if the absolute value of the wave vector $k$ is smaller than $\sigma_{CME}$. From (\ref{anomalous_maxwell_dispersion}) it is also easy to see that for a given wave vector $\vec{k}$, only one of two helical mode will exhibit exponential growth. For example, for $\mu_A > 0$ (and hence $Q_A > 0$ and $\sigma_{CME} > 0$) and $\sigma_{CME } > k > 0$ the exponentially growing solution has the form
\begin{eqnarray}
\label{exp_growing_solution}
 E_1 = f e^{\kappa t} \cos\lr{k x_3}, \,
 E_2 = - f e^{\kappa t} \sin\lr{k x_3},
 \nonumber \\
 B_1 = -f \frac{k}{\kappa} e^{\kappa t} \cos\lr{k x_3}, \,
 B_2 =  f \frac{k}{\kappa} e^{\kappa t} \sin\lr{k x_3} ,
 \nonumber \\
 E_3 = 0, \, B_3 = 0,
\end{eqnarray}
where $f$ is an arbitrary amplitude and $\kappa \equiv -i w = -\frac{\sigma}{2} + \sqrt{\frac{\sigma^2}{4} - k^2 + \sigma_{CME} k}$. It is important to stress that since this solution grows monotonously in time, here we use the terms ``circular polarization'', ``handedness'' and ``helicity'' to describe the rotation of the vectors  $\vec{E}$ and $\vec{B}$ along the $x_3$ axis, rather than in time. The growth of long-wavelength electromagnetic waves and the decay of short-wavelength waves predicted by the anomalous Maxwell equations (\ref{anomalous_maxwell}) is a novel mechanism for the inverse cascade in relativistic magnetohydrodynamics \cite{Boyarsky:12:1,Boyarsky:15:1}, which transfers energy from long- to short-wavelength helical magnetic fields.

 The fact that the exponentially growing solution (\ref{exp_growing_solution}) has the helical structure of the form (\ref{exp_growing_solution}) also suggests the mechanism which can stop the growth of electromagnetic field at later times. Namely, let us recall that for massless chiral fermions the time evolution of the axial charge is governed by the anomaly equation:
\begin{eqnarray}
\label{volume_integrated_anomaly}
 \partial_t Q_A = \frac{g^2}{2 \pi^2} \int d^3x \vec{E} \cdot \vec{B} ,
\end{eqnarray}
where the axial charge $Q_A = Q_R - Q_L$ is defined as the difference between the charges $Q_R$ and $Q_L$ of the right- and left-handed fermions, $g$ is the electromagnetic coupling constant and we have integrated over space to get rid of the spatial divergence of the axial current. For simplicity, in this paper we consider only a single flavor of Dirac fermions with electromagnetic coupling $g = 1$.

 For the exponentially growing solution (\ref{exp_growing_solution}) the product $\vec{E} \cdot \vec{B}$ is negative: $\vec{E} \cdot \vec{B} = -f^2 \frac{k}{\kappa} e^{2 \kappa t}$ \footnote{It is interesting that nonzero scalar product $\vec{E} \cdot \vec{B}$ is only possible for exponentially growing or decaying solutions. E.g. circularly polarized electromagnetic waves in a dissipationless medium always have $\vec{E} \cdot \vec{B} = 0$.}. The anomaly equation (\ref{volume_integrated_anomaly}) then dictates that the time derivative $\partial_t Q_A$ of the axial charge is negative. Since we have assumed $Q_A > 0$, $\mu_A > 0$, we see that the growing helical solution (\ref{exp_growing_solution}) will result in the decrease of $Q_A$ and hence of $\mu_A$. This depletion of chirality imbalance should eventually suppress the chiral magnetic conductivity in (\ref{cme_current}) and hence slow down or stop completely the exponential growth in (\ref{exp_growing_solution}).

 However, the above analysis of the chiral plasma instability, which follows \cite{Boyarsky:12:1,Tashiro:12:1,Yamamoto:13:1,Sadofyev:13:1,TorresRincon:13:1,TorresRincon:14:1,Tuchin:14:1,Kharzeev:15:1,Shovkovy:16:1}, essentially relies on an assumption that the electric current takes the form $\vec{j} = \sigma \vec{E} + \sigma_{CME} \vec{B}$ with constant ohmic and chiral magnetic conductivities. In reality, both $\sigma$ and $\sigma_{CME}$ depend on the frequency and wave vector of electromagnetic field in a nontrivial way \cite{Kharzeev:09:1,Ren:11:1,Yee:14:1,Buividovich:13:8,Landsteiner:11:2,Landsteiner:12:1,Buividovich:15:1}. One can also expect a strong dependence of $\sigma$ and $\sigma_{CME}$ on the spatial and temporal modulation of the axial charge density, which will in general appear at late evolution times \cite{Shovkovy:16:1}. Moreover, as the instability might lead to quite large strengths of electric and magnetic fields, nonlinear effects beyond the linear response result (\ref{cme_current}) might become important. Using linear response approximation to describe the interactions between the fermions and the electromagnetic fields is in fact similar to the Lyapunov analysis of the full quantum evolution, which is in general nonlinear. What concerns inter-fermion interactions, so far they were taken into account only indirectly, by using the relaxation time approximation \cite{Manuel:15:1} or the decoherence of the fermionic wave functions \cite{Sadofyev:13:2}.
A consistent inclusion of all these effects in the anomalous Maxwell equations (\ref{anomalous_maxwell}) would be certainly difficult with approaches based e.g. on the chiral kinetic theory \cite{Stephanov:12:1,TorresRincon:13:1,TorresRincon:14:1,Shovkovy:16:1}, chiral hydrodynamics \cite{Son:09:1,Boyarsky:15:1} or the Langevin-type effective theory \cite{Rothkopf:15:1}.

 This situation clearly calls for a more first-principle description of the real-time dynamics of chirally imbalanced plasma which would overcome these limitations and approximations.  In this paper, we report on the numerical study of the real-time chiral plasma instability within the framework of the so-called classical statistical field theory (CSFT) \cite{Aarts:98:1,Aarts:00:1,Borsanyi:09:1,Tanji:13:1,Berges:13:1,Berges:14:1}, which captures the first nontrivial order of the expansion of the full quantum evolution operator in powers of the Planck constant. CSFT is currently the state-of-the art method for numerical simulations of real-time quantum evolution. The CSFT approximation is justifiable as long as the characteristic occupation numbers of the physically relevant gauge field modes are large. That is, the dynamics of the gauge fields should be almost classical. On the other hand, the real-time dynamics of fermions is exact in CSFT. Taking into account that in all previous studies gauge fields were also treated classically, the applicability of the CSFT approach is obviously wider than that of the previously used approaches. An approach very similar to CSFT has been recently used in \cite{Fukushima:15:1} to study the real-time dynamics of CME. However, in this work the back-reaction of fermions on the electromagnetic field, which is the origin of the chiral plasma instability, was not taken into account. The real-time dynamics of axial charge was also studied in $1 + 1$-dimensional Abelian Higgs model in the pioneering work \cite{Aarts:98:1}.

 Our studies are based on the non-compact formulation of lattice quantum electrodynamics, which avoids potential problems with monopole condensation in the strong-coupling phase \cite{Polyakov:75:1}. For fermions, we use the massless Wilson-Dirac Hamiltonian which has a low-energy chiral symmetry. At sufficiently high energies, this symmetry is broken due to the Wilson term. In the condensed matter context, this breaking is a natural feature of any model description of Dirac and Weyl semimetals \cite{Vazifeh:13:1,Hosur:13:1,Sekine:13:1}. In Section~\ref{sec:chirality_pumping} we demonstrate that the effect of this explicit breaking is, however, not very large (see also \cite{Berges:16:1}). Therefore we hope that our results should be also at least qualitatively relevant in the context of high-energy physics, where the chiral symmetry is exact at the level of the Lagrangian, or tends to be exact at sufficiently high energies.

\begin{figure}
\includegraphics[width=3cm]{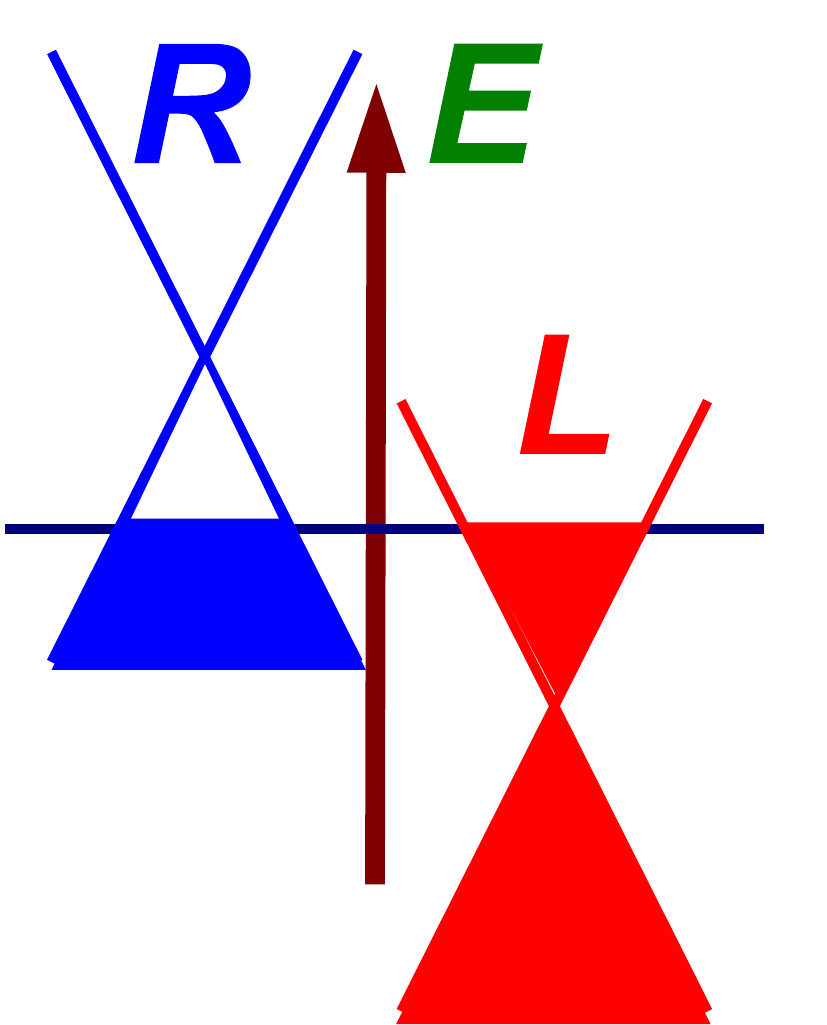}
\includegraphics[width=3cm]{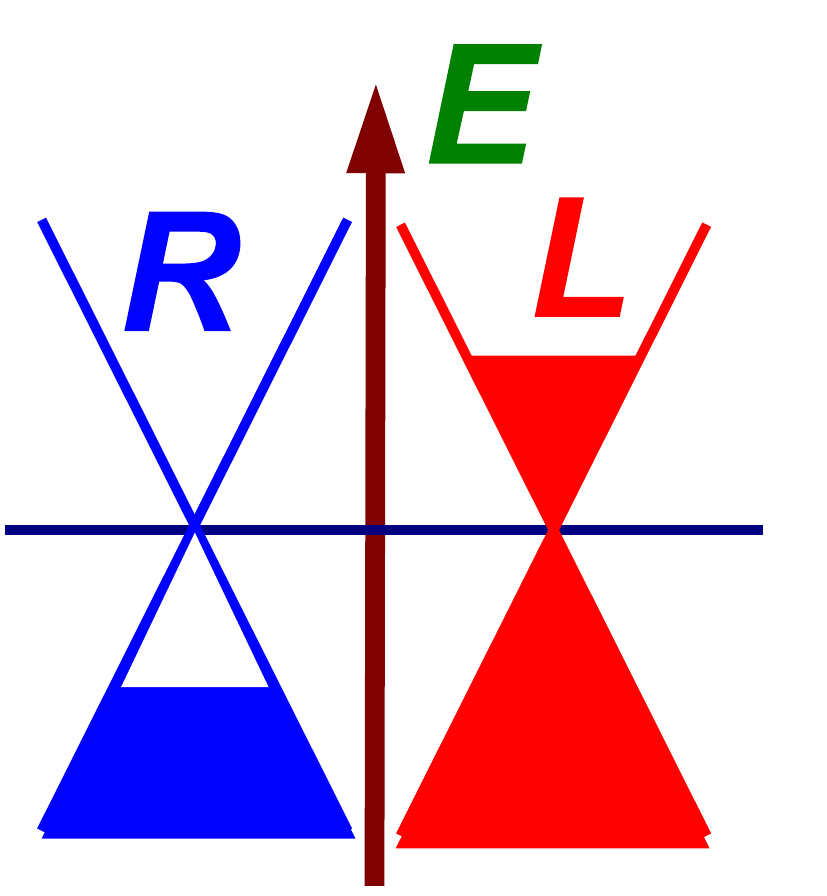}\\
\caption{Two ways of introducing initial chiral imbalance for the many-body Dirac Hamiltonian. On the left: by introducing the chiral chemical potential $\mu_A$ in the single-particle Dirac Hamiltonian. On the right: by filling more right-handed eigenstates and less left-handed eigenstates (or vice versa).}
\label{fig:init_mua_methods}
\end{figure}

 In order to introduce the initial chirality imbalance, we have started the simulations with a state in which more right-handed eigenstates and less left-handed eigenstates are filled, as depicted on the right panel of Fig.~\ref{fig:init_mua_methods}. Such a state is an excited state of the many-body Dirac Hamiltonian, even in the absence of electromagnetic fields. It is an idealized description of the result of ``chirality pumping'' process in parallel electric and magnetic fields \cite{Kharzeev:14:1,Parameswaran:13:1} or in intense circularly polarized laser beams.

 Alternatively, we have also considered the introduction of the chiral chemical potential into the single-particle Dirac Hamiltonian, which changes the energies of the right- and the left-handed Dirac points (see left panel of Fig.~\ref{fig:init_mua_methods}). Such initial state has nonzero axial charge but is still the ground state of the many-body Hamiltonian in the absence of interactions with electromagnetic fields. In simulations which started from this state we have not found any signatures of instability or the transfer of helicity from fermions to electromagnetic fields. The axial charge density exhibited only small fluctuations on top of the large mean value. Presumably, the reason for such a behavior is that nonzero chiral chemical potential corresponds to the physical situation in which our system is connected to an infinite reservoir of axial charge, which is capable of maintaining its initial value at a constant level. Since the anomaly equation (\ref{volume_integrated_anomaly}) holds also at nonzero chiral chemical potential, this also implies that the magnetic helicity can only exhibit small fluctuations, possibly related to the violation of the anomaly equation due to lattice artifacts. Since such behavior is not really interesting, we do not discuss this setup in what follows.

 The structure of this paper is the following: in Section \ref{sec:methods} we start with a brief summary of the details of our numerical CSFT algorithm. In Appendix \ref{apdx:csft_derivation} we provide a more detailed derivation of this algorithm with a bias towards non-relativistic field theories and condensed matter systems, which might hopefully complement the existing literature on CSFT (see e.g. \cite{Aarts:98:1,Berges:14:1} for derivations which are more in the spirit of relativistic quantum field theory). In this Appendix we also demonstrate explicitly the absence of any nontrivial Jacobian in the integration measure in the CSFT algorithm, and discuss some practical aspects of our CSFT simulations on parallel computers. In Section \ref{sec:chirality_pumping} we present the results of the simulations of the chirality pumping process. First, we consider chirality pumping in external parallel electric and magnetic fields in the absence of backreaction and verify the validity of the anomaly equation (\ref{volume_integrated_anomaly}) in our numerical setup. After that we consider the effect of backreaction of fermionic current on the chirality pumping process and demonstrate that the dynamical screening of the external electric field prevents the system from acquiring large axial charge density at late evolution times.

 In Section \ref{sec:chiral_instability}, we consider the decay of the initial chiral imbalance and the generation of electromagnetic fields with nonzero helicity. In order to trigger the decay, we start simulations with several initially excited modes of electromagnetic field. Following the energies of helical modes in momentum space, we demonstrate that only long-wavelength modes of definite helicity grow and all other modes decay with time. This is a direct numerical evidence of the inverse cascade phenomenon \cite{Boyarsky:12:1,Boyarsky:15:1,Kharzeev:15:1,Yamamoto:16:1} due to chiral plasma instability. We find, however, that the dependence of the strength of the inverse cascade on the initial conditions and parameters of the simulations is significantly more complex than predicted by the anomalous Maxwell equations (\ref{anomalous_maxwell}). In particular, in the simulations which exhibit most rapid growth of helical magnetic fields the axial charge does not decay at all. Correspondingly, the mechanism which stops the inverse cascade in our simulations is not related to the axial charge decay, again in contrast to the expectations based on the equations (\ref{anomalous_maxwell}) \cite{Boyarsky:12:1,Boyarsky:15:1,Tashiro:12:1,Yamamoto:13:1,Manuel:15:1,Kharzeev:15:1}. Rather, we observe that in most simulations which do exhibit axial charge decay the inverse cascade emerges for the electric rather than for magnetic fields. In Section \ref{sec:conclusions} we conclude with a general discussion of our results and an outlook.

\section{Classical statistical field theory approximation to real-time evolution}
\label{sec:methods}

 We consider the many-body fermionic Hamiltonian coupled to dynamical non-compact electromagnetic fields on the lattice, so that the full Hamiltonian $\hat{H}$ of our system is $\hat{H} = \hat{H}_F + \hat{H}_{EM}$. The fermionic Hamiltonian $\hat{H}_F$ reads
\begin{eqnarray}
\label{hferm_def}
 \hat{H}_F = \sum\limits_{x, y} \hat{\psi}^{\dag}_x h_{x,y} \hat{\psi}_y ,
\end{eqnarray}
where the labels $x$, $y$ denote the sites of the three-dimensional cubic lattice, $\hat{\psi}^{\dag}_x$, $\hat{\psi}_x$ are the spinor-valued fermionic creation and annihilation operators which satisfy the anti-commutation relation $\lrc{\hat{\psi}^{\dag}_x, \hat{\psi}_y} = \delta_{xy}$ and $h_{x,y}$ is the massless single-particle Wilson-Dirac Hamiltonian with the Wilson coefficient $r = 1$:
\begin{eqnarray}
\label{h_wd}
h_{x,y} = 3 v_F \, \beta \delta_{x,y}
 +
\frac{i \, v_F}{2}  \sum_{i=1}^3
  \lr{i \beta + \alpha_i} e^{ i g A_{    x,i}} \delta_{y,x+e_i}
  + \nonumber\\ +
\frac{i \, v_F}{2}  \sum_{i=1}^3
  \lr{i \beta - \alpha_i} e^{-i g A_{x-e_i,i}} \delta_{y,x-e_i}  .
\end{eqnarray}
Here $A_{x,i}$ is the vector potential of the lattice gauge field, $e_i$ denotes the unit lattice vector in the direction $i$, $v_F$ is the Fermi velocity, $\beta$ and $\alpha_i$ are the Dirac $\beta$- and $\alpha$-matrices and $\gamma_5$ is the generator of chiral rotations:
\begin{eqnarray}
\label{DiracMatricesDef}
 \beta = \left(
 \begin{array}{cc}
 0 & 1  \\
 1 & 0  \\
 \end{array}
 \right), \,
 \alpha_i = \left(
 \begin{array}{cc}
 \sigma_i & 0  \\
 0 &  -\sigma_i  \\
 \end{array}
 \right), \,
 \gamma_5 = \left(
 \begin{array}{cc}
 1 &  0  \\
 0 &  -1  \\
 \end{array}
 \right) ,
\end{eqnarray}
where $\sigma_i$ are the Pauli matrices. In (\ref{h_wd}) we have assumed that the lattice spacing is unity. Thus in what follows all dimensionful quantities are expressed in units of the lattice spacing.

 The lattice Hamiltonian $\hat{H}_{EM}$ of electromagnetic field is the straightforward lattice discretization of the corresponding continuum Hamiltonian:
\begin{eqnarray}
\label{h_em}
\hat{H}_{EM} = \sum_x \sum_{i=1}^3  \lr{ \frac{\hat{E}_{x,i}^2}{2} +\sum_{j=i}^3 \frac{\hat{F}_{x,ij}^2}{2} + \hat{A}_{x,i} \mathcal{J}_{x, i}\lr{t} } ,
\end{eqnarray}
where $\mathcal{J}_{x, i}\lr{t}$ is the external current (which is required, e.g., to switch the external electric and magnetic fields on and off) and the operator of the magnetic field strength tensor $\hat{F}_{x,ij}$ is defined in terms of the finite differences of the vector potential operator $\hat{A}_{x,i}$ as
\begin{eqnarray}
\label{f_def}
 \hat{F}_{x,ij} = \hat{A}_{x,i} + \hat{A}_{x+e_i,j} - \hat{A}_{x+e_j,i} - \hat{A}_{x,j} .
\end{eqnarray}
The operators of the electric field $\hat{E}_{x,i}$ and the vector potential $\hat{A}_{x,i}$ are canonically conjugate and satisfy the commutation relations ${\lrs{\hat{E}_{x,i}, \hat{A}_{y,j}} = -i \delta_{xy} \delta_{ij}}$. We impose periodic boundary conditions in all spatial directions both for the gauge and the fermionic fields.

 In our CSFT algorithm, described in detail in Appendix \ref{apdx:csft_derivation}, we numerically solve the classical equations of motion of the electromagnetic field with the Hamiltonian (\ref{h_em})
\begin{eqnarray}
\label{continuum_maxwell}
 \partial_t^2 A_{x,i}\lr{t}
 =
 - \mathcal{J}_{x,i}\lr{t}
 -
 \vev{j_{x,i}\lr{t}}
 - \nonumber \\ -
 \sum_j \lr{F_{x,ij}\lr{t} - F_{x-e_j,ij}\lr{t}} ,
\end{eqnarray}
where the initial value of the time derivative $\left. \partial_t A_{x,i}\lr{t} \right|_{t = 0}$ is the initial value of the electric field $E_{x,i}\lr{0}$ and $\vev{j_{x,i}\lr{t}}$ is the vacuum expectation value of the fermionic electric current, which can be calculated as
\begin{eqnarray}
\label{fermionic_current_vev}
 \vev{j_{x,i}\lr{t}}
 = \tr\left(
 \rho_0 \,
 u\lr{0, t} \,
 j_{x,i} \,
 u^{\dag}\lr{0, t}
 \right) ,
\end{eqnarray}
where $j_{x,i} = \frac{\partial h}{\partial A_{x,i}}$ is the single-particle operator of electric current, $u\lr{0, t}$ is the quantum evolution operator defined by the single-particle Schr\"{o}dinger equation
\begin{eqnarray}
\label{single_particle_schrodinger}
 \partial_t u\lr{0, t} = i h\lrs{A_{x,i}\lr{t}} u\lr{0, t}, \quad u\lr{0, 0} = I ,
\end{eqnarray}
and $\rho_0$ is the initial density matrix which characterizes the initial occupation numbers $n_a$ of single-particle states $\ket{\psi_a}$:
\begin{eqnarray}
\label{single_particle_density_matrix}
 \rho_0 = \sum\limits_a \ket{\psi_a} n_a \bra{\psi_a} .
\end{eqnarray}
In our case, $\ket{\psi_a}$ are the eigenstates of the single-particle Hamiltonian (\ref{h_wd}). If some occupation numbers are exactly zero (which can be the case at zero temperature), some components of the quantum evolution operator $u$ completely decouple and can be discarded in the solution of the equation (\ref{single_particle_schrodinger}). This can be used to speed up the algorithm, typically by a factor of two (for a standard zero-temperature Fermi distribution). The expectation value $\vev{j_{x,i}\lr{t}}$ in (\ref{continuum_maxwell}) describes the effect of backreaction of fermions on the electromagnetic fields.

 We thus have a closed set of equations (\ref{continuum_maxwell}), (\ref{fermionic_current_vev}) and (\ref{single_particle_schrodinger}), which allows to evolve the fermionic quantum states and the classical electromagnetic fields in a self-consistent way. One can also explicitly check that this evolution conserves the total energy $H_{EM} + \vev{\hat{H}_F}$ of electromagnetic field and fermions up to the work done by the external current $\mathcal{J}^{x,i}\lr{t}$:
\begin{eqnarray}
\label{energy_conservation}
 \partial_t \lr{H_{EM} + \vev{\hat{H}_F}} = - \sum\limits_{x,i} \mathcal{J}_{x,i}\lr{t} E_{x,i}\lr{t} ,
 \nonumber \\
 H_{EM} = \frac{1}{2}\sum\limits_{x,i}
 \lr{\lr{\partial_t A_{x,i}}^2
 + \sum\limits_j F_{x,ij}^2 } ,
 \nonumber \\
 \vev{\hat{H}_F} = \tr\left(
 \rho_0 \,
 u\lr{0, t} \,
 h \,
 u^{\dag}\lr{0, t}
 \right) .
\end{eqnarray}
We have solved the evolution equations (\ref{continuum_maxwell}) and (\ref{single_particle_schrodinger}) using the leap-frog integrator, which slightly violates the conservation of energy (\ref{energy_conservation}). At sufficiently small time step this violation is completely under numerical control, see Fig.~\ref{fig:energy_joint} in the Appendix \ref{apdx:csft_derivation}.

 In the CSFT approach, one can also partially take into account the quantum fluctuations of the electromagnetic fields, encoded in the nontrivial Wigner transform $\bar{\rho}_{EM}\lr{A_0, E_0}$ of the initial density matrix $\rho_{EM}$, where $A_0 \equiv A\lr{t=0}$ and $E_0 \equiv E\lr{t=0}$ are the initial values of electric and magnetic fields. To this end one should additionally average all observables over $A_0$ and $E_0$, sampled with the probability $\bar{\rho}_{EM}\lr{A_0, E_0}$\footnote{This of course requires that the initial density matrix should correspond to a sufficiently classical state, so that the Wigner transform of the initial density matrix is non-negative.}. However, in our work we have not taken these initial quantum fluctuations into account for the following reasons. First, in the case of anomaly equation (\ref{volume_integrated_anomaly}) with massless fermions we have found that the effect of initial fluctuations of electromagnetic fields is much more significant than in the case of e.g. Schwinger pair production \cite{Berges:14:1}\footnote{In the process of development of our algorithm we have reproduced the results of \cite{Berges:14:1} on the Schwinger pair production with massive fermions and explicitly checked, that initial quantum fluctuations of electromagnetic field encoded in the nontrivial initial density matrix $\hat{\rho}_{EM} \sim e^{-\hat{H}_{EM}/T}$ with $T \rightarrow 0$ (so that only the ground state energies $\hbar w/2$ contribute) have little effect on the Schwinger pair production rate. Correspondingly, only a few samples of $A_0$ and $E_0$ are enough for reliably computing the expectation values.}, and hence much more samples of the initial fields are required to reach acceptable statistical errors. Partially this can be explained by the large value of the electromagnetic coupling constant, which was $g = 1.0$ in most of our simulations. We expect that the role of initial fluctuations will be smaller for a smaller value of $g$, say, $g = 0.1$. In the latter case, however, the characteristic time scale of the chiral plasma instability increases significantly above our current simulation times.

 In addition, taking into account initial quantum fluctuations of electromagnetic fields makes it impossible to assume spatial homogeneity of electromagnetic fields along some of the lattice directions, which is essential to speed up the CSFT simulations at large lattice sizes.

 Thus while the effect of quantum fluctuations on the chiral plasma instability might be potentially very significant and interesting, we cannot study it with our presently available computational resources and leave it for the future work. In this work, we avoid the statistical averaging over the initial values of the fields $E_0$, $A_0$ by using the very simple form of the initial density matrix $\bar{\rho}_{EM}\lr{A_0, E_0}$ which is just a delta-function on some particular, specifically chosen initial values. We thus completely neglect the quantum fluctuations of the electromagnetic fields. Nevertheless, this approximation is still certainly wider than the chiral kinetic theory or hydrodynamical approximation.

\section{Chirality pumping in parallel electric and magnetic fields}
\label{sec:chirality_pumping}

 In this Section we study the real-time evolution of the axial charge $Q_A$ in the background of constant parallel external electric and magnetic fields. In the absence of backreaction, such setup provides a direct check of how well the anomaly equation (\ref{volume_integrated_anomaly}) holds for the Wilson-Dirac Hamiltonian with inexact chiral symmetry \cite{Berges:16:1}, which we further use to study the chiral plasma instability in Section~\ref{sec:chiral_instability}. The effect of backreaction is also interesting since the anomaly equation (\ref{volume_integrated_anomaly}) is known to receive nontrivial corrections if the electromagnetic fields are dynamical \cite{Adler:04:1,Anselm:89:1,Jensen:13:1}.

 In order to induce the constant external electric field $\vec{E} = E \, \vec{e}_3$, we switch on the external current of the form $\mathcal{J}_{x,i}\lr{t} = \delta_{i,3} \, E \, t$. Constant external magnetic field is induced by the static circular external current flowing around the plaquettes with $x_1 = L_1 - 1$, $x_2 = L_2 - 1$ for all $x_3 = 0 \ldots L_3 - 1$, where $L_1$, $L_2$ and $L_3$ are lattice sizes. This static current is like a thin solenoid piercing a stack of lattice plaquettes, with the field strength being equal to $B$ outside of solenoid and $B - B L_1 L_2$ inside it. Lattice fermions, however, acquire only the Aharonov-Bohm phase $e^{i g B}$ when encircling such a solenoid if one imposes the flux quantization condition
\begin{eqnarray}
\label{flux_quantization}
 g B L_1 L_2 = 2 \pi \Phi, \quad \Phi \in \mathbb{Z} .
\end{eqnarray}
The total external current which we insert in the equations (\ref{continuum_maxwell}) is the sum of the two currents which create constant electric and magnetic fields.

 For the initial state of the fermionic fields, we use the eigenstates of the Wilson-Dirac Hamiltonian $h\lrs{A_0}$, where $A_0$ is the initial gauge field configuration with constant magnetic field $B$ as described above. In this work we consider only the limit of zero temperature, correspondingly, only the states with negative energies are initially occupied.

 The Wilson-Dirac Hamiltonian which we use in our simulations does not have exact chiral symmetry, and there is no uniquely defined axial charge operator which would exactly satisfy the anomaly equation (\ref{volume_integrated_anomaly}) and commute with the Hamiltonian. Rather, the anomaly equation (\ref{volume_integrated_anomaly}) can only hold approximately, in the limit of large lattice volume and sufficiently smooth, slowly changing and small gauge fields \cite{Karsten:81:1,Rothe:98:1}. We thus use the simplest possible definitions of the operators of the axial charge density $q_{A \, x}$ and the total axial charge $Q_A$:
\begin{eqnarray}
\label{axial_charge_def}
 \hat{q}_{A \, x} = \hat{\psi}^{\dag}_x \gamma_5 \hat{\psi}_x, \quad
 \hat{Q}_A = \sum\limits_x \hat{q}_{A \, x}.
\end{eqnarray}
The time-dependent expectation value of the axial charge density is calculated similarly to the expectation value of electric current in (\ref{fermionic_current_vev}):
\begin{eqnarray}
\label{axial_charge_vev}
 \vev{q_{A \, x}\lr{t}}
 = \tr\left(
 \rho_0 \,
 u\lr{0, t} \,
 \gamma_5 \, P_x \,
 u^{\dag}\lr{0, t}
 \right) ,
\end{eqnarray}
where $P_x$ is the single-particle projector on a single lattice site $x$: $\lrs{P_x}_{x_1 x_2} = \delta_{x_1 x} \delta_{x x_2}$.

\begin{figure*}[h!tpb]
  \centering
  \includegraphics[width=0.5\textwidth]{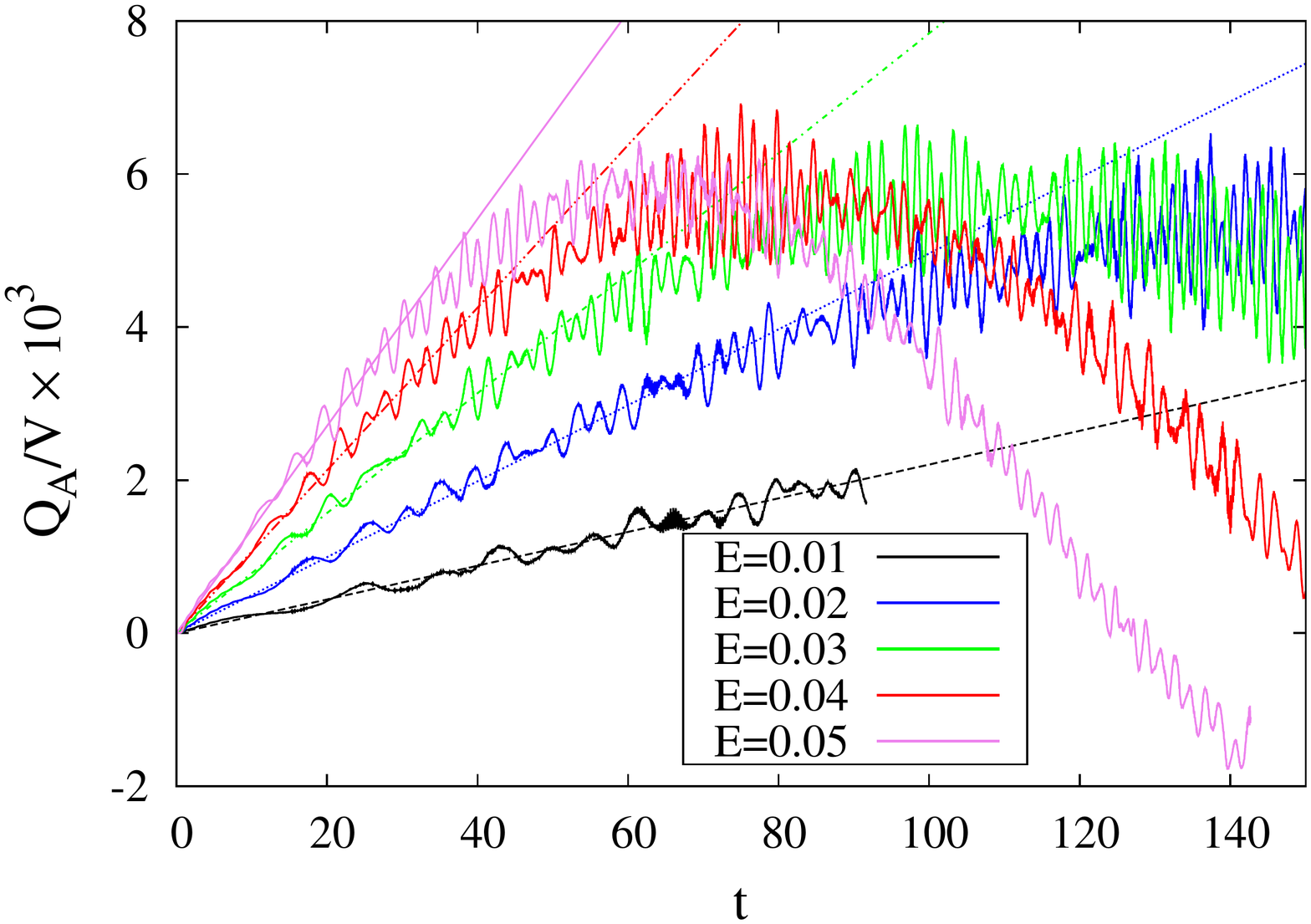}\includegraphics[width=0.5\textwidth]{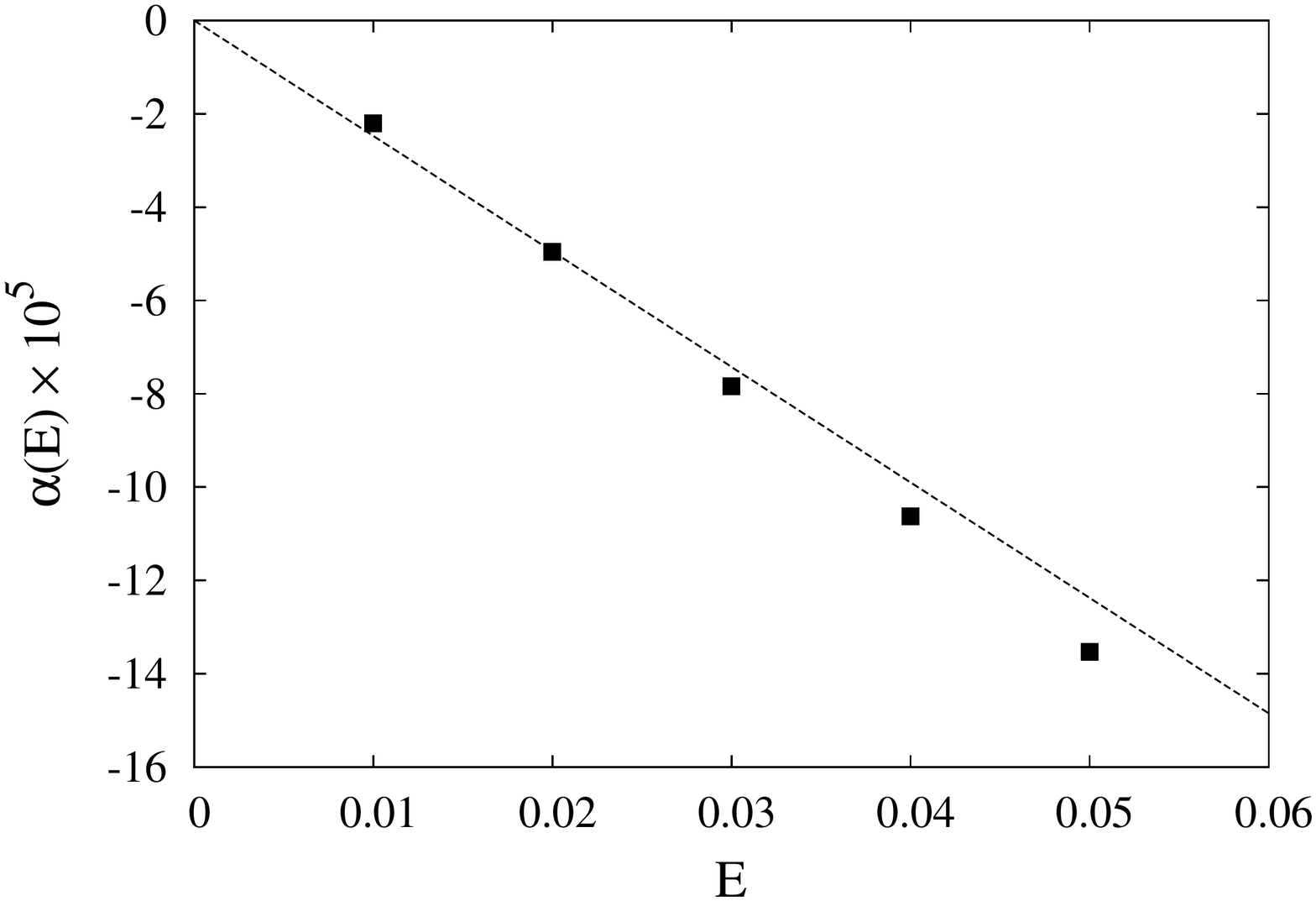}\\
  \caption{Chirality pumping without backreaction for various external electric fields and external magnetic field with flux $\Phi = 1$ on the $10 \times 10 \times 32$ lattice. On the left: time dependence of the axial charge $Q_A$ and its linear fits at early times. On the right: dependence of the slope of these fits on the external electric field and the linear fit of this dependence. Here and in what follows time is expressed in units of lattice spacing.}
  \label{fig:E_field_dep}
\end{figure*}

 First we neglect the backreaction of fermionic electric current on the electromagnetic field and measure the time dependence of the axial charge in constant parallel external electric and magnetic fields. The results are shown on the left panel of Fig.~\ref{fig:E_field_dep} for the $10 \times 10 \times 32$ lattice with $\Phi = 1$ quantum of magnetic field flux. One can see that $Q_A$ grows linearly with time until it reaches some maximal value $Q_A/V \approx 0.006$, where $V = L_1 L_2 L_3$ is the total number of lattice sites (lattice volume). After that, $Q_A$ decreases again. This decrease is a lattice artifact related to the fact that the characteristic momentum $p \sim E t$ of fermions accelerated by an electric field $E$ approaches the UV cutoff set by the compact size of the lattice momentum space $k_{\mu} \in \lrs{-\pi \ldots \pi}$. Due to the periodicity of lattice momentum space, at large time scales the behavior of the axial charge (in the absence of backreaction) is well described by $Q_A \sim \sin\lr{E t/2}$. There are also some short-time fluctuations on top of the clearly visible linear growth at early times.

\begin{figure}[h!tpb]
  \centering
  \includegraphics[width=0.45\textwidth]{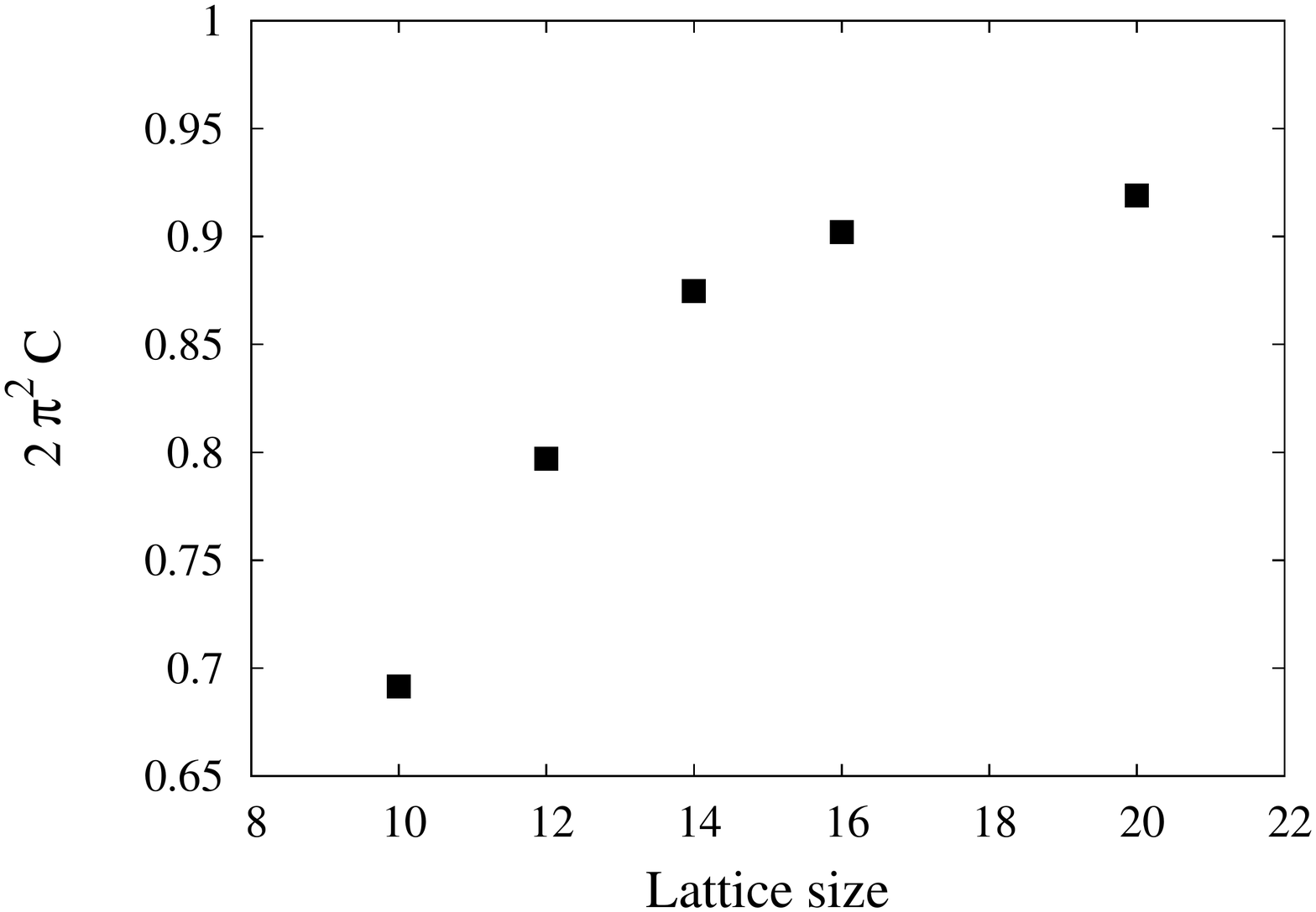}\\
  \caption{Dependence of the lattice anomaly coefficient ($\partial_t Q_A = C \int d^3\vec{x} \vec {E} \cdot \vec{B}$) on the transverse lattice size. Lattice size in the direction parallel to the magnetic field is fixed to $L_3 = 32$.}
  \label{fig:anomaly_size}
\end{figure}

 In order to estimate the linear growth rate at early times, we perform the linear fit of the form $Q_A\lr{t}/V = \alpha\lr{E} \, t$ in the range $t \in \lrs{0 \ldots 50}$ for $E = 0.01$ and $E = 0.02$ and in the range $t \in \lrs{0 \ldots 30}$ for other values of $E$. The dependence of the coefficient $\alpha\lr{E}$ on the electric field is shown on the right panel of Fig.~\ref{fig:E_field_dep}. Again, this dependence is linear with a good precision, and we perform another linear fit $\alpha\lr{E} = C E$, where $C$ corresponds to the anomaly coefficient relating $\partial_t Q_A$ and $\int d^3\vec{x} \vec{E} \cdot \vec{B}$ in (\ref{volume_integrated_anomaly}). On Fig.~\ref{fig:anomaly_size} we show the dependence of $C$ on the size of the lattice (in the directions perpendicular to the magnetic field). One can see how $C$ approaches the value $C = \frac{1}{2 \pi^2}$ in the limit of large lattices, in agreement with the anomaly equation (\ref{volume_integrated_anomaly}). Let us also note that for larger number of flux quanta one can perform a similar fitting procedure. However, finite-volume artifacts in $C$ are significantly larger for larger magnetic fluxes. For this reason, in this work we have only used external magnetic field with one flux quantum.

\begin{figure*}[h!tpb]
  \centering
  \includegraphics[width=0.5\textwidth]{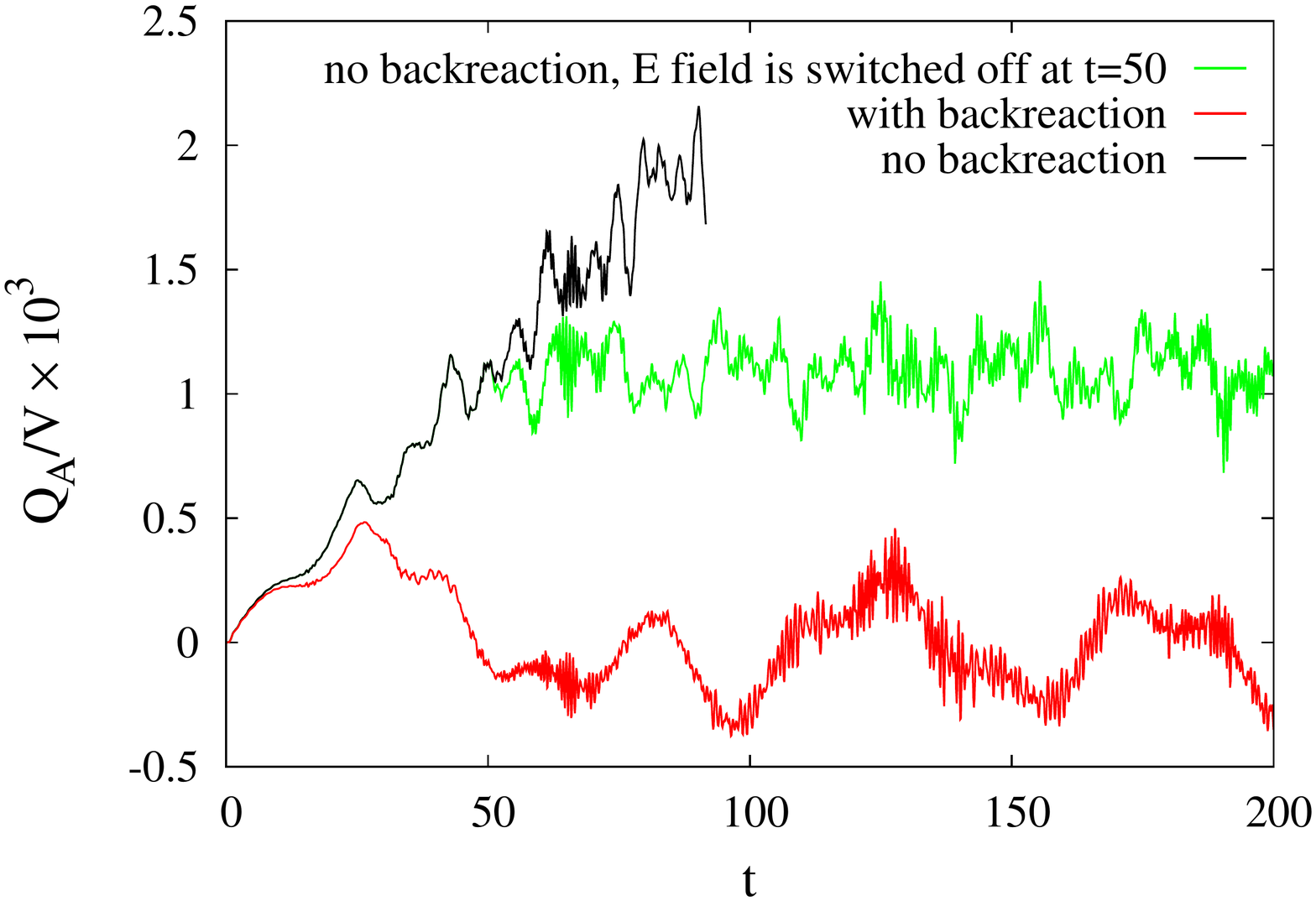}\includegraphics[width=0.5\textwidth]{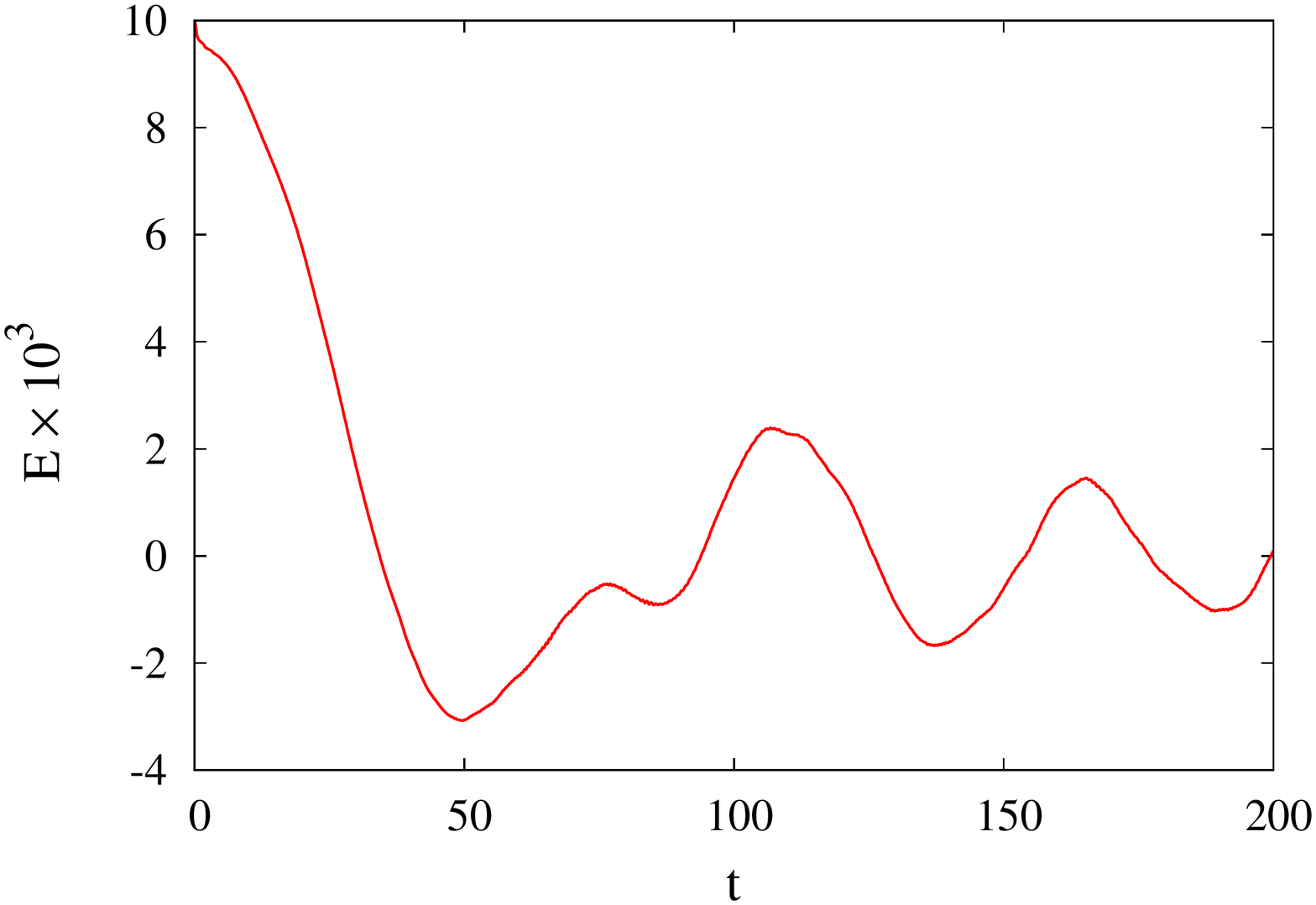}\\
  \caption{A comparison of chirality pumping processes with and without backreaction of the fermionic electric current on the electromagnetic field. On the left: time dependence of the axial charge. On the right: time dependence of the component of the volume-averaged electric field parallel to the magnetic field.  Lattice size is $10\times 10 \times 32$, the flux of external magnetic field is $\Phi = 1$, external electric field is $0.01$.}
  \label{fig:effect_BR}
\end{figure*}

 It is also interesting to check how the axial charge depends on time after the external electric field is switched off and the external magnetic field remains constant (which we believe to be a more realistic experimental setup than the simultaneous switching off of all fields). The time dependence of the axial charge for such a situation is shown on the left plot of Fig.~\ref{fig:effect_BR} (green line). External electric field is switched off at the time $t = 50$. One can see that starting from this moment of time the axial charge exhibits only some small-scale fluctuations around the nonzero mean value. This demonstrates that the effect of explicit chiral symmetry breaking due to the Wilson term in the Hamiltonian (\ref{h_wd}) is rather small for such simulation parameters, and the total axial charge is almost a conserved quantity.

  After establishing the validity of the anomaly equation (\ref{volume_integrated_anomaly}) in our simulation setup, we study the effect of backreaction of dynamical electromagnetic fields on the chirality pumping process. Technically, the backreaction is taken into account by inserting the expectation value of the fermionic electric current $\vev{j_{x,i}}$ into the Maxwell equations for the electromagnetic field. We now consider the situation in which the external electric and magnetic fields are switched on permanently. As we will see, in simulations with backreaction switching off the electric field at sufficiently late times anyway does not affect the evolution significantly due to screening by the dynamically generated electric field. Time dependence of the axial charge $Q_A\lr{t}$ for simulation with backreaction is shown on Fig.~\ref{fig:effect_BR}. For comparison, on the same Figure we also show $Q_A\lr{t}$ for simulations without backreaction, where the electric field is permanent or switched off at $t = 50$.

 One can see that while at $t \lesssim 30$ $Q_A\lr{t}$ grows approximately linearly with $t$ both with and without backreaction, at later times backreaction leads to a rapid decay of $Q_A$ with subsequent fluctuations around zero. In order to understand the origin of this effect, remember that axial anomaly can be also regarded as the Schwinger pair production in the effective $1 + 1$-dimensional theory of fermions on the lowest Landau level. It is thus natural to expect that particle-anti-particle pairs produced by the external electric field will tend to screen this field, just as in the case of Schwinger effect in $\lr{3+1}$ dimensions. To check this conjecture, on the right panel of Fig.~\ref{fig:effect_BR} we plot the volume-averaged electric field projected on the direction of the magnetic field. One can see that indeed it quite quickly decreases from the initial value $E = 0.01$, reaching zero at around $t \approx 30$ - exactly at the time at which the growth of the axial charge stops (see left panel of the same Figure). After that the electric field exhibits some fluctuations around zero with the amplitude which is approximately five times smaller than the initial field value. We thus conclude that the effect of backreaction on the chirality pumping is to stop the growth of the axial charge by screening the external electric field down to zero.

\section{Chiral plasma instability and decay of axial charge}
\label{sec:chiral_instability}

 In this Section, we consider a situation in which some initial chiral imbalance is already created e.g. by chirality pumping, and the parallel electric and magnetic fields are adiabatically switched off while keeping nonzero value of the total axial charge $Q_A$ and hence the chiral chemical potential $\mu_A$. In this setup we would like to study the existence and the late-time evolution of the exponentially growing solutions (\ref{exp_growing_solution}) of the anomalous Maxwell equations (\ref{anomalous_maxwell}), as well as the associated inverse cascade of energy of helical electromagnetic fields.

 In order to implement the initial chirality imbalance as discussed in the introductory Section \ref{sec:intro} (see right panel of Fig.~\ref{fig:init_mua_methods}), we divide all the eigenstates of the single-particle Wilson-Dirac Hamiltonian $h\lrs{A_0}$, where $A_0$ is the initial value of the vector potential, into the positive chirality states with $\bra{\psi_a} \gamma_5 \ket{\psi_a} > 0$ and the negative chirality states with  $\bra{\psi_a} \gamma_5 \ket{\psi_a} < 0$. For positive chirality states we fill all the levels with $\epsilon_a < \mu_A$, and for negative chirality states - all the levels with $\epsilon_a < - \mu_A$. While the eigenstates of the Wilson-Dirac Hamiltonian are not in general the eigenstates of the $\gamma_5$ operator, for eigenstates with sufficiently small momenta our definition is maximally close to the notion of distinct Fermi levels of left- and right-handed continuum massless fermions. In practice, this definition is unambiguous as long as all the energy levels $\epsilon_a$ are non-degenerate. In the case of degenerate energy levels (as e.g. in the case of the Wilson-Dirac Hamiltonian with zero gauge fields or in the background of a single plane wave), one can additionally rotate the eigenstates within the degenerate subspaces in order to maximize the absolute values of matrix elements $\bra{\psi_a} \gamma_5 \ket{\psi_a}$.

\begin{figure*}[h!tpb]
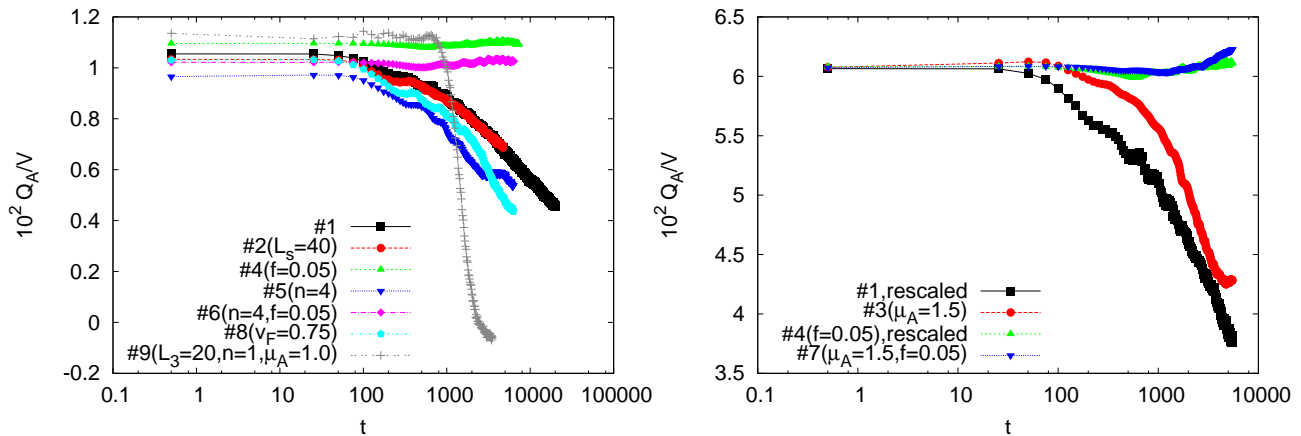

  \centering
  \includegraphics[width=6cm,angle=-90]{{{q5_vs_t_mu0.75}}}\includegraphics[width=6cm,angle=-90]{{{q5_vs_t_mu1.50}}}\\
  \caption{Time dependence of the total axial charge for different simulation parameters. On the right plot we compare simulations with different $\mu_A$ and linearly rescale $q_A\lr{t} \rightarrow c q_A\lr{t}$ so that the initial values of $c q_A\lr{0}$ agree for all simulations. In the plot labels, numbers preceded by the hash symbol \# correspond to the numbers of parameter sets in Table \ref{tab:parameter_sets}. In parentheses we give the values of only those parameters which differ from the default parameters ($L_3=200$, $L_s=20$, $\mu_A=0.75$, $n=10$, $f=0.2$ and $v_F = 1$, parameter set No.~1).}
\label{fig:q5_vs_t}
\end{figure*}

\begin{figure}[h!tpb]
  \centering
  \includegraphics[width=6cm,angle=-90]{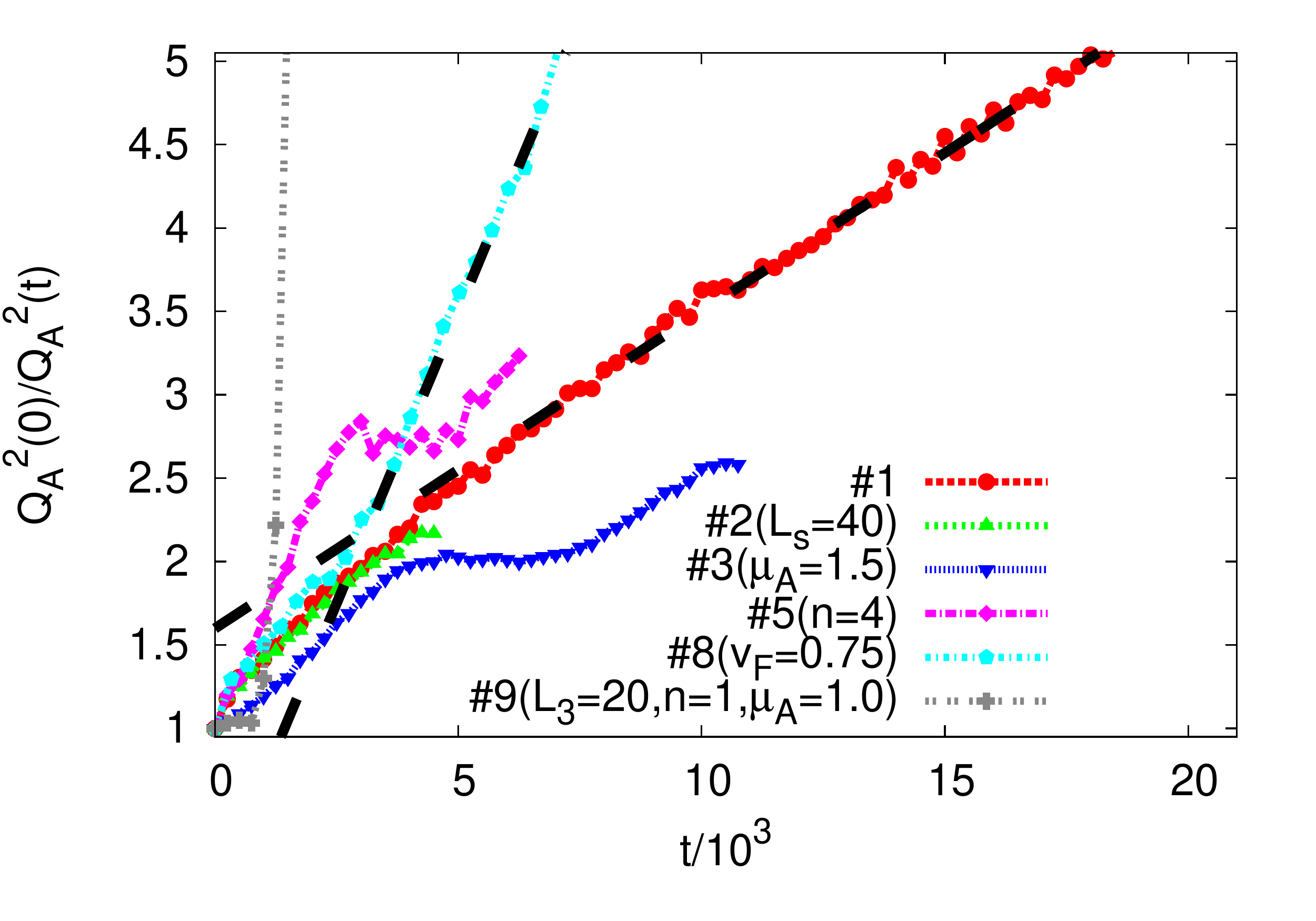}\\
  \caption{Universal late-time scaling $Q_A\lr{t} \sim 1/\sqrt{t}$ of axial charge density in simulations with $f = 0.2$. Time dependence of $\lr{Q_A\lr{0}/Q_A\lr{t}}^2$ in the second half of the total evolution time is fitted, where appropriate, by linear functions $\lr{Q_A\lr{0}/Q_A\lr{t}}^2 = A + B t$. The fits are shown with dashed black lines.}
\label{fig:q5_vs_t_scaling}
\end{figure}

\begin{figure*}
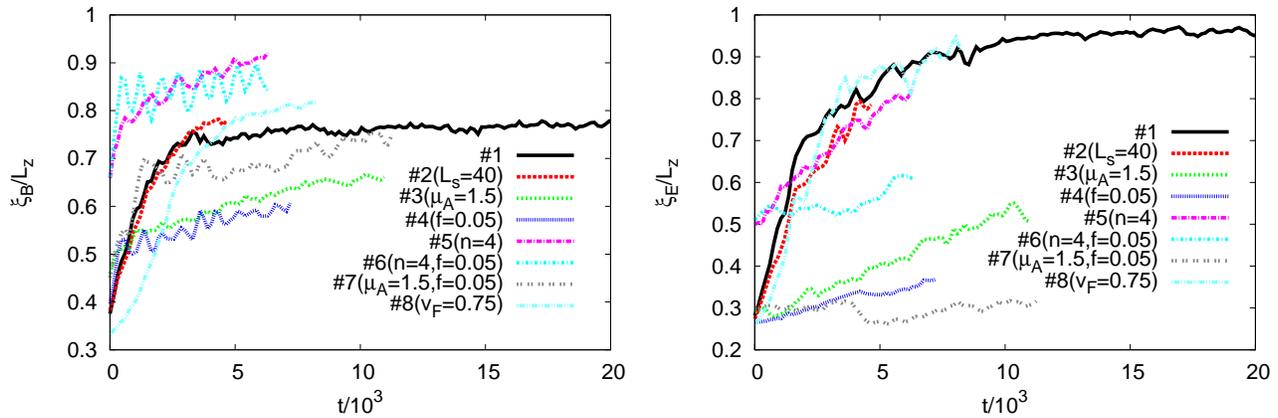

  \centering
  \includegraphics[width=6cm,angle=-90]{{{xB_vs_t}}}\includegraphics[width=6cm,angle=-90]{{{xE_vs_t}}}\\
  \caption{Magnetic and electric correlation lengths $\xi_B$ and $\xi_E$ (defined in (\ref{elmag_corr_def})) as functions of time for different parameter sets. A smearing procedure similar to (\ref{smeared_energies_def}) was applied in order to suppress minor short-scale fluctuations in the data.}
  \label{fig:xEB_vs_t}
\end{figure*}

\begin{figure}[h!tpb]
  \centering
  \includegraphics[width=6cm,angle=-90]{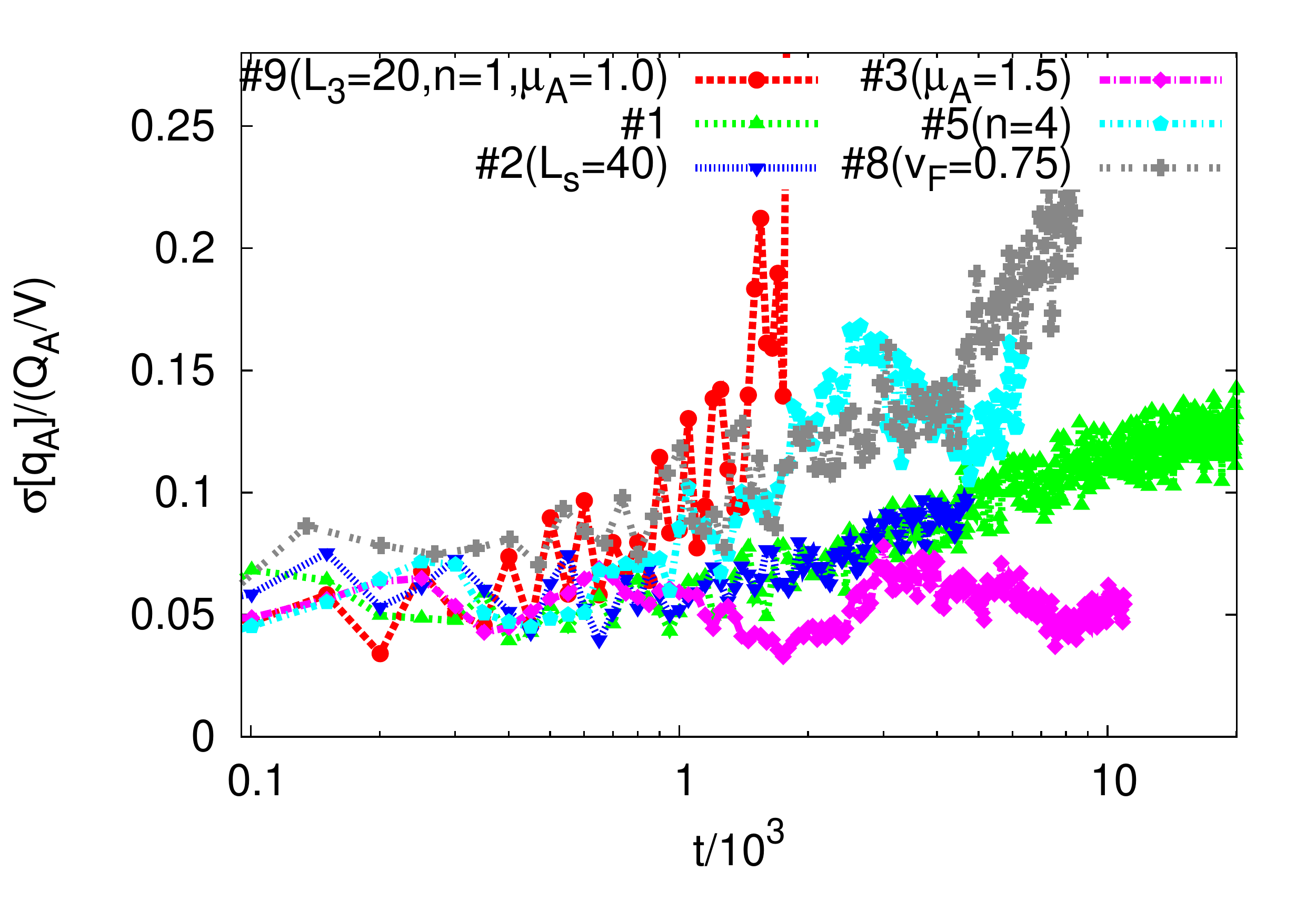}\\
  \caption{Time dependence of the standard deviation of the axial charge density $\sigma\lrs{q_A}$ from its volume-averaged value $Q_A/V$ in simulations with $f = 0.2$.}
\label{fig:q5dis_vs_t}
\end{figure}

\begin{figure*}
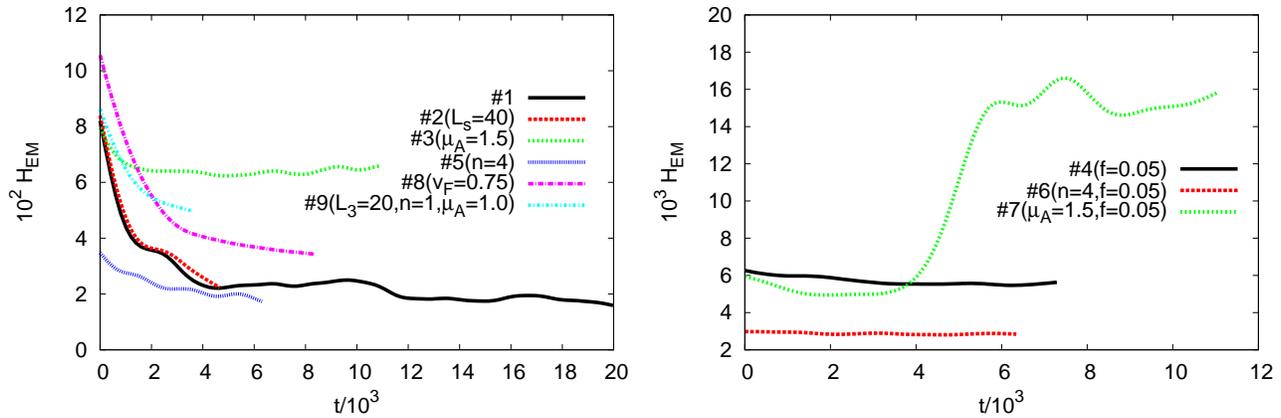

  \centering
  \includegraphics[width=6cm,angle=-90]{{{energies_vs_t_f0.2}}}\includegraphics[width=6cm,angle=-90]{{{energies_vs_t_f0.05}}}\\
  \caption{Time dependence of the total energy of electromagnetic field in simulations with $f = 0.2$ (on the left) and $f = 0.05$ (on the right). A smearing procedure similar to (\ref{smeared_energies_def}) was applied in order to suppress minor short-scale fluctuations in the data.}
  \label{fig:em_energy_vs_t}
\end{figure*}

 With lattice discretizations of the Dirac Hamiltonian it is not possible to have very large values of the chiral chemical potential $\mu_A$, since the dispersion relation at the Fermi energy $\mu_A \gtrsim 1$ starts deviating from the Dirac cone $\epsilon\lr{\vec{k}} = v_F |\vec{k}|$ due to lattice artifacts. At $\mu_A = 2$, the Fermi energy touches the lowest van Hove singularity (saddle point) of the dispersion relation, and the excitations around the Fermi surface no longer correspond to Dirac fermions. On the other hand, the solution (\ref{exp_growing_solution}) of the anomalous Maxwell equations (\ref{anomalous_maxwell}) with the conventional value $\sigma_{CME} = \frac{\mu_A}{2 \pi^2}$ of the chiral magnetic conductivity suggests that the wave vectors at which the chiral plasma instability can occur are bounded by $|\vec{k}| < \frac{\mu_A}{2 \pi^2}$. On a finite spatial lattice of size $L$ with periodic boundary conditions, the smallest nonzero value of $|\vec{k}|$ is $|\vec{k}| = \frac{2 \pi}{L}$, which dictates the lower bound on the size of the lattice where the instability can be observed:
\begin{eqnarray}
\label{length_bound_estimate}
 L > \frac{4 \pi^3}{\mu_A} .
\end{eqnarray}
Thus it is advantageous to use large values of $\mu_A$ in order to reduce the lattice size used for simulations. Taking the moderate value $\mu_A = 0.75$, at which the dispersion relation is still linear with a good precision, we obtain $L > 165$. Performing simulations on an isotropic three-dimensional lattice of such a size would be a formidable numerical task. For this reason we have used the lattices with different sizes in different directions, so that the size $L_3$ in the direction $x_3$ of electromagnetic wave propagation is much larger than the sizes $L_1 = L_2 \equiv L_s$ in the transverse directions $x_1$ and $x_2$. In addition, we have assumed that electromagnetic fields do not depend on the transverse coordinates $x_1$ and $x_2$. This allows us to represent the single-particle evolution operator $u\lr{0, t}$ in the block-diagonal form in the basis of plane waves propagating along $x_1$ and $x_2$, which greatly reduces the dimensionality of the linear space on which the single-particle Schr\"{o}dinger equation (\ref{single_particle_schrodinger}) should be solved. By comparing the results of simulations with $L_s = 20$ and $L_s = 40$ at fixed $L_3=200$ (see Table \ref{tab:parameter_sets} and Figs.~\ref{fig:q5_vs_t}, \ref{fig:q5_vs_t_scaling}, \ref{fig:xEB_vs_t}, \ref{fig:q5dis_vs_t} and \ref{fig:em_energy_vs_t}) we have checked that the dependence on the transverse lattice size is rather weak. Let us also note that one of the reasons for not using the final state of chirality pumping process described in Section \ref{sec:chirality_pumping} for the study of chiral plasma instability is that in this case it is not possible to assume spatial homogeneity in transverse directions due to the breaking of translational invariance by external magnetic field \cite{Wiese:08:1}.

 While the fermionic initial state described above is an excited state which can spontaneously decay due to chiral plasma instability, in numerical simulations one always needs some small ``seed'' perturbation to start the decay process in a controllable way. For this reason we have started our simulations with a state in which also some finite number $n$ of electromagnetic field modes are excited. All of them are plane waves propagating along the lattice direction $x_3$ with the largest size $L_3$, with a few smallest nonzero wave numbers $k_m = \frac{2 \pi m}{L_3}$, $m = 1 \ldots n$ and random linear polarizations. In order to facilitate the detection of the inverse cascade, we choose the amplitudes of all modes in such a way that their contributions to the total energy of electromagnetic field are equal. Thus the explicit form of our initial electromagnetic field configuration is
\begin{eqnarray}
\label{initial_plane_wave}
 A_{x,i}\lr{t = 0} = \sum\limits_{m=1}^{n} \frac{f}{w\lr{k_m}} n_{m \, i} \cos\lr{k_m x_3 + \phi_m},
 \nonumber \\
 E_{x,i}\lr{t = 0} \equiv \partial_t \left.A_{x,i}\lr{t}\right|_{t=0}
 = \nonumber \\ =
 \sum\limits_{m=1}^{n} f \, n_{m \, i} \, \sin\lr{k_m x_3 + \phi_m},
\end{eqnarray}
where $n_{m \, i}$ are the random unit transverse polarization vectors which are chosen to coincide with one of the basis vectors $\vec{e}_1$, $\vec{e}_2$ with equal probability, $\phi_m \in \lrs{0, 2 \pi}$ are the random phases and $w\lr{k_m} = \sqrt{4 \sin^2\lr{\frac{k_m}{2}}}$ corresponds to the lattice dispersion relation for free massless fields on the lattice.

\begin{table}
 \centering
 \begin{tabular}{|c|c|c|c|c|c|c|c|c|c|}
  \hline
  Set.No. & $L_3$ & $L_s$ & $\mu_A$ & $n$ & $f$ & $v_F$ & $Q_A\downarrow$ & $I^B_k\uparrow$ & $I^E_k\uparrow$\\
  \hline
  1 & 200 & 20 & 0.75 & 10 & 0.20  & 1.00 & \cmark & \xmark & \cmark \\
  2 & 200 & 40 & 0.75 & 10 & 0.20  & 1.00 & \cmark & \xmark & \cmark \\
  3 & 200 & 20 & 1.50 & 10 & 0.20  & 1.00 & \cmark & \cmark & \cmark \\
  4 & 200 & 20 & 0.75 & 10 & 0.05  & 1.00 & \xmark & \cmark & \xmark \\
  5 & 200 & 20 & 0.75 &  4 & 0.20  & 1.00 & \cmark & \cmark & \qmark \\
  6 & 200 & 20 & 0.75 &  4 & 0.05  & 1.00 & \xmark & \cmark & \qmark \\
  7 & 200 & 20 & 1.50 & 10 & 0.05  & 1.00 & \xmark & \cmark & \qmark \\
  8 & 200 & 20 & 0.75 & 10 & 0.20  & 0.75 & \cmark & \xmark & \cmark \\
  9 &  20 & 20 & 1.00 &  1 & 0.20  & 1.00 & \cmark & \xmark & \xmark \\
  \hline
 \end{tabular}
 \caption{Summary of parameters and results of our simulations of chiral plasma instability. The column $Q_A\downarrow$ summarizes the decay of the axial charge and the columns $I^{E,B}_k\uparrow$ summarize the growth of the energies of long-wavelength electric and magnetic fields. The symbols \cmark, \xmark \, and \qmark \, denote, respectively, the clearly visible growth, clearly visible absence of growth and intermediate situations for which it is difficult to make any conclusion within a finite simulation time.}
 \label{tab:parameter_sets}
\end{table}

 In order to understand how the evolution process depends on various lattice parameters, we have performed simulations with $9$ different parameter sets, which are summarized in Table \ref{tab:parameter_sets}. We have varied both the transverse and the longitudinal lattice sizes, the initial electromagnetic field amplitude $f$, the number $n$ of initially excited electromagnetic field modes, the initial value of axial charge and the Fermi velocity $v_F$. The parameter set No.~1 with $L_3=200$, $L_s=20$, $\mu_A=0.75$, $n=10$, $f=0.2$ and $v_F = 1$ is the ``default'' parameter set, and all other sets differ from it by a change in a few parameters. Correspondingly, in what follows we label the data points on the plots which combine the results from several simulations by the number of parameter set (preceded by the hash symbol \#), in parentheses giving only those parameters which differ from the default ones.

 On Fig.~\ref{fig:q5_vs_t} we show the time dependence of axial charge in simulations with parameters summarized in Table \ref{tab:parameter_sets}. In simulations with initial amplitude of electromagnetic fields being equal to $f = 0.2$ the axial charge $Q_A$ decays with time. Interestingly, simulations with the smallest lattice size (parameter set No.~9) exhibit the fastest decay of $Q_A$. On the other hand, with the initial amplitude $f = 0.05$ the axial charge density exhibits only a rather small decrease at intermediate evolution times, subsequently followed by a slight increase. This nontrivial dependence on the electromagnetic field strength suggests that the dynamics of the decay process is more complicated than suggested by the anomalous Maxwell equations (\ref{anomalous_maxwell}). It is interesting that the evolution of the axial charge seems to depend only weakly on simulation parameters other than the initial amplitude $f$ and the longitudinal lattice size $L_3$ (through the value of the lowest wave number $\frac{2 \pi}{L_3}$). Even the dependence on the chiral chemical potential $\mu_A$ appears to be rather weak (after a trivial rescaling with respect to the initial value), see right plot on Fig.~\ref{fig:q5_vs_t}. The characteristic time scale for the evolution of the axial charge appear to be essentially larger than in the chirality pumping simulations in the previous Section. This difference can be qualitatively explained by much weaker field strengths in the simulations described in this Section. We also note that the initial values of the axial charge are roughly consistent with the continuum formula $Q_A/V = \mu_A^3/\lr{3 \pi^2}$, where $V$ is the lattice volume. Deviations from this value can be explained, first, by the inclusion of the initial vector potential $A_0$ in the initial Hamiltonian, and second, by the smaller value of chirality $\left|\bra{\psi_a} \gamma_5 \ket{\psi_a}\right| < 1$ for high-energy eigenstates $\ket{\psi_a}$ of the Wilson-Dirac Hamiltonian (\ref{h_wd}).

 A scaling analysis of the anomalous Maxwell equations suggests that at late evolution times the time dependence of the axial charge density approaches the simple power law \cite{Kharzeev:15:1,Yamamoto:16:1}
\begin{eqnarray}
\label{universal_scaling}
 Q_A\lr{t} \sim 1/\sqrt{t} .
\end{eqnarray}
In order to check this scaling, on Fig.~\ref{fig:q5_vs_t_scaling} we plot the time dependence of the inverse square of the axial charge, which should approach the linear function $1/Q_A^2\lr{t} \sim t$ according to (\ref{universal_scaling}). This asymptotic behavior indeed seems to emerge at late evolution times for simulations with $L_s = 20$, $n = 10$, $f = 0.2$ and $\mu_A = 0.75$, both with $v_F = 1$ and $v_F = 0.75$ (parameter sets No.~1~and~8). The linear fits of $\lr{Q_A\lr{0}/Q_A\lr{t}}^2$ for these simulations are shown on Fig.~\ref{fig:q5_vs_t_scaling} with dashed black lines.

 We now check whether the decay of the axial charge is accompanied by the growth of the long-wavelength modes of electromagnetic field, as predicted by the anomalous Maxwell equations (\ref{anomalous_maxwell}). To this end we perform the Fourier transforms of the transverse electric and magnetic fields (taking into account that they depend only on the $x_3$ coordinate)
\begin{eqnarray}
\label{el_field_fourier}
 E_{k,i}\lr{t} = \frac{1}{\sqrt{L_3}} \sum\limits_{x_3} e^{i k x_3} E_{x,i}\lr{t} ,
 \nonumber \\
 B_{k,i}\lr{t} = \frac{1}{\sqrt{L_3}} \sum\limits_{x_3} e^{i k x_3} B_{x,i}\lr{t} ,
\end{eqnarray}
where $i = 1, 2$, and further decompose the Fourier-transformed fields into the helical components $E_{k,R/L}\lr{t}$ and $B_{k,R/L}$ with right- and left-handed helicities:
\begin{eqnarray}
\label{circular_polarizations}
 B_{k,R}\lr{t} = \frac{1}{2}\lr{B_{k,1}\lr{t} + B_{-k,1}\lr{t}}
 + \nonumber \\ +
 \frac{1}{2 i}\lr{B_{k,2}\lr{t} - B_{-k,2}\lr{t}} ,
 \nonumber \\
 B_{k,L}\lr{t} = \frac{1}{2 i}\lr{B_{k,1}\lr{t} - B_{-k,1}\lr{t}}
 + \nonumber \\ +
 \frac{1}{2}\lr{B_{k,2}\lr{t} + B_{-k,2}\lr{t}} .
\end{eqnarray}
For electric fields, the definition of helical components is exactly the same. Again, here the term ``helicity'' refers to the direction of rotation of transverse electric and magnetic fields along the spatial direction of wave propagation (the $x_3$ axis in our setup).

 After such a decomposition, we calculate the energies of left- and right-handed helical electric and magnetic fields with a given wave number $k$ as
\begin{eqnarray}
\label{circular_energies}
 I^B_{k,R/L}\lr{t} = \left|B_{k,R/L}\lr{t}\right|^2/2 + \left|B_{-k,R/L}\lr{t}\right|^2/2 ,
 \nonumber \\
 I^E_{k,R/L}\lr{t} = \left|E_{k,R/L}\lr{t}\right|^2/2 + \left|E_{-k,R/L}\lr{t}\right|^2/2 ,
\end{eqnarray}
where $k = \frac{2 \pi m}{L_3}$ and $m = 0 \ldots \floor{L_3/2}$ now spans only the half of the discrete lattice momenta. Since the initial configuration of electromagnetic fields contains plane waves with (random) linear polarizations and equal energies, at $t = 0$ the energies $I_{k,R/L}\lr{t}$ of all left- and right-handed electromagnetic modes with $0 < k \leq \frac{2 \pi n}{L_3}$ are equal.

\begin{figure}
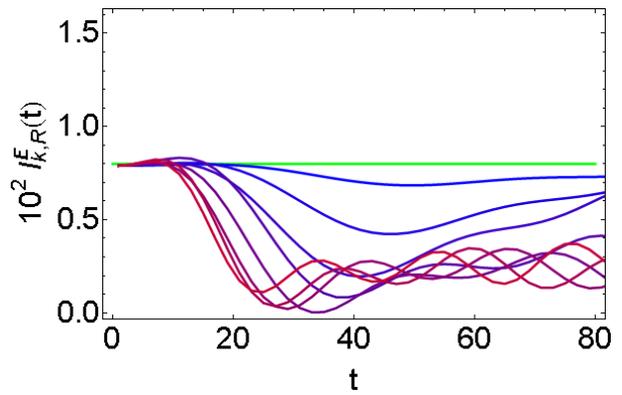

  \centering
  \includegraphics[width=8cm]{{{Lx20_10modes_mu=0.75_f=0.2_Enlarged}}}\\
  \caption{Time dependence of the energies $I^E_{k,R/L}\lr{t}$ of right-handed components of electric field on a short time interval at the beginning of evolution for parameter set No.~1 ($L_s = 200$, $n = 10$, $\mu_A = 0.75$, $f = 0.2$). The wave numbers are coded in color, from pure blue for the smallest nonzero value $k = \frac{2 \pi}{L_3}$ (largest wavelength) to pure red for $k = \frac{2 \pi n}{L_3}$.}
  \label{fig:spectrum_enlarged}
\end{figure}

 We have found that for all simulations the energies $I^{E,B}_{k,R/L}\lr{t}$ exhibit quite large short-scale fluctuations with period of order $\Delta t \sim 10 \ldots 100$, which is smaller for short-wavelength modes and larger for long-wavelength ones. For illustration, see Fig.~\ref{fig:spectrum_enlarged}, where we plot the time dependence of $I^E_{k,R}\lr{t}$ within a short initial period of time for simulation with parameter set No.~1. These oscillations indicate that the helical magnetic and electric fields represented by the basis (\ref{circular_polarizations}) are not the eigenstates of the evolution process, which is in sharp contrast to the solution (\ref{exp_growing_solution}) of the anomalous Maxwell equations (\ref{anomalous_maxwell}). We have explicitly checked that if the backreaction of fermions on the electromagnetic fields is neglected, these oscillations disappear and the energies $I^{E,B}_{k,R/L}\lr{t}$ are constant in time for all values of $k$ and for all polarizations. This observation suggests that the short-scale oscillations might originate from the nontrivial dependence of fermionic current on the frequency, wave number and amplitude of electromagnetic field, which turns the solutions of the anomalous Maxwell equations (\ref{anomalous_maxwell}) into waves with generic elliptic polarizations.

 Despite the short-scale fluctuations, we still find it useful to decompose our fields in the basis (\ref{circular_polarizations}), since the corresponding electromagnetic modes carry definite helicity and thus the energies of helical modes can be used to define, at least approximately, the helicity on the lattice. This definition is advantageous since direct lattice discretizations of the continuum formula $\mathcal{H} \sim \int d^3x \vec{A} \cdot \vec{B}$ are in general flawed by lattice artifacts. In order to abstract ourselves from the short-scale fluctuations, we define the energies $\bar{I}_{k,R/L}\lr{t}$ which are averaged over some finite time interval $T$:
\begin{eqnarray}
\label{smeared_energies_def}
 \bar{I}^{E,B}_{k,R/L}\lr{t} = \frac{1}{T} \int\limits_{t-T/2}^{t+T/2} dt' I^{E,B}_{k,R/L}\lr{t'} .
\end{eqnarray}
We have used the value $T = 25$, which is sufficient to remove practically all short-scale oscillations.

\begin{figure*}[p!]
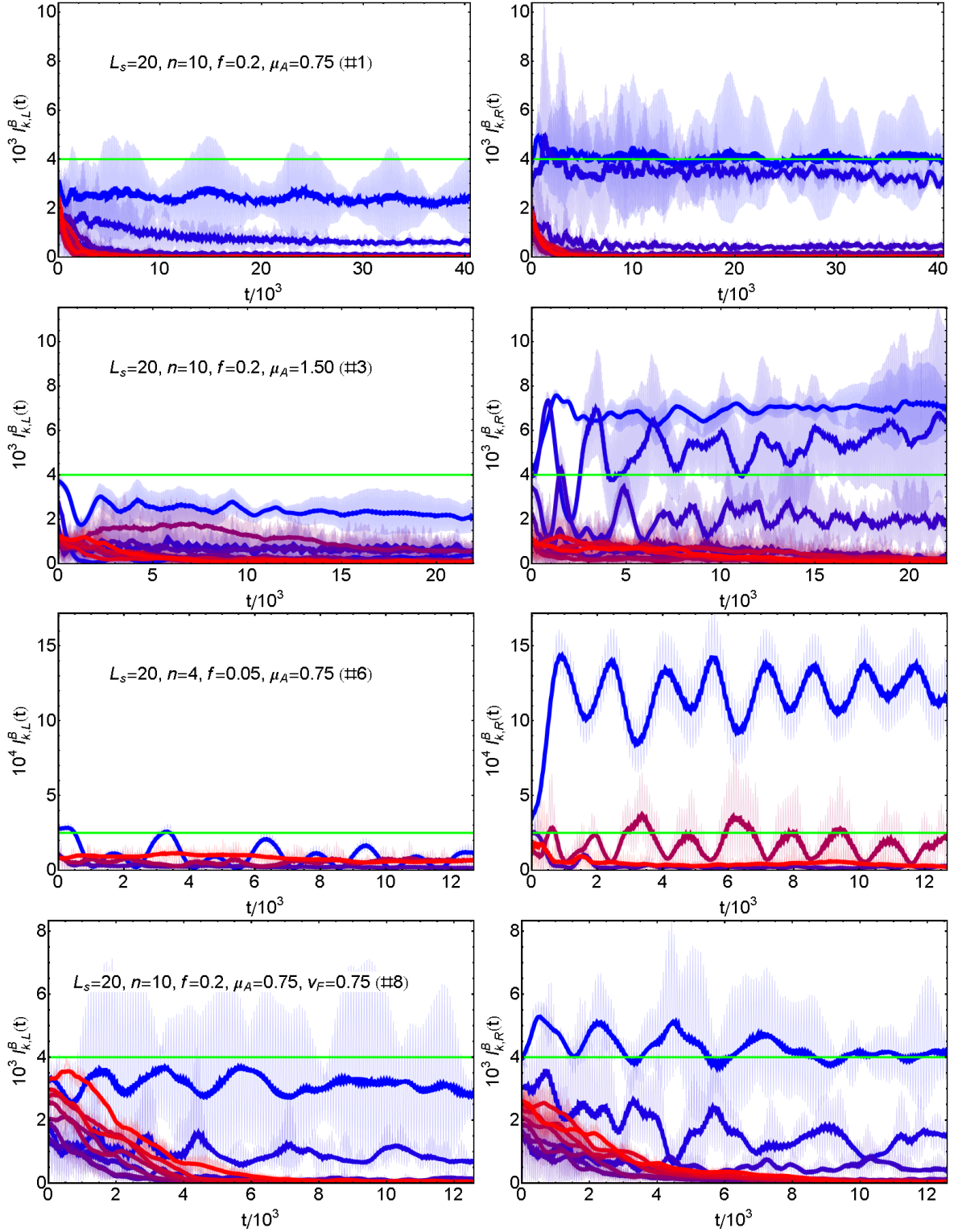

  \centering
  \includegraphics[width=0.43\textwidth]{{{Lx20_10modes_mu=0.75_f=0.2_LB}}}
  \includegraphics[width=0.43\textwidth]{{{Lx20_10modes_mu=0.75_f=0.2_RB}}}\\
  \includegraphics[width=0.43\textwidth]{{{x200_s20_n10_f0.2000_m1.5000_LB}}}
  \includegraphics[width=0.43\textwidth]{{{x200_s20_n10_f0.2000_m1.5000_RB}}}\\
  \includegraphics[width=0.43\textwidth]{{{x200_s20_n4_f0.0500_m0.7500_LB}}}
  \includegraphics[width=0.43\textwidth]{{{x200_s20_n4_f0.0500_m0.7500_RB}}}\\
  \includegraphics[width=0.43\textwidth]{{{Lx20_10modes_mu=0.75_f=0.2_vf0.75_LB}}}
  \includegraphics[width=0.43\textwidth]{{{Lx20_10modes_mu=0.75_f=0.2_vf0.75_RB}}}\\
  \caption{Time dependence of the energies of helical \textbf{magnetic fields} for several selected sets of simulation parameters. The wave numbers are coded in color, from pure blue for the smallest nonzero value $k = \frac{2 \pi}{L_3}$ (largest wavelength) to pure red for $k = \frac{2 \pi n}{L_3}$. Semi-transparent colored regions show the range of short-scale oscillations of $I^{B}_{k,R/L}\lr{t}$, and thick solid lines show the time dependence of the time-smeared energies $\bar{I}^B_{k,R/L}\lr{t}$ defined in (\ref{smeared_energies_def}). In black-and-white version pure blue and pure red correspond to black and light grey, respectively. Horizontal green (light grey) lines show the initial energies which are equal for all modes. Left-handed and right-handed modes are in the left and in the right columns, respectively.}
  \label{fig:B_power_spectra}
\end{figure*}

\begin{figure*}[p!]
  \centering
  \includegraphics[width=0.43\textwidth]{{{Lx20_10modes_mu=0.75_f=0.2_LE}}}
  \includegraphics[width=0.43\textwidth]{{{Lx20_10modes_mu=0.75_f=0.2_RE}}}\\
  \includegraphics[width=0.43\textwidth]{{{x200_s20_n10_f0.2000_m1.5000_LE}}}
  \includegraphics[width=0.43\textwidth]{{{x200_s20_n10_f0.2000_m1.5000_RE}}}\\
  \includegraphics[width=0.43\textwidth]{{{x200_s20_n4_f0.0500_m0.7500_LE}}}
  \includegraphics[width=0.43\textwidth]{{{x200_s20_n4_f0.0500_m0.7500_RE}}}\\
  \includegraphics[width=0.43\textwidth]{{{Lx20_10modes_mu=0.75_f=0.2_vf0.75_LE}}}
  \includegraphics[width=0.43\textwidth]{{{Lx20_10modes_mu=0.75_f=0.2_vf0.75_RE}}}\\
  \caption{Time dependence of the energies of helical \textbf{electric fields} for several selected sets of simulation parameters. The wave numbers are coded in color, from pure blue for the smallest nonzero value $k = \frac{2 \pi}{L_3}$ (largest wavelength) to pure red for $k = \frac{2 \pi n}{L_3}$. Semi-transparent colored regions show the range of short-scale oscillations of $I^{E}_{k,R/L}\lr{t}$, and thick solid lines show the time dependence of the time-smeared energies $\bar{I}^E_{k,R/L}\lr{t}$ defined in (\ref{smeared_energies_def}). In black-and-white version pure blue and pure red correspond to black and light grey, respectively. Horizontal green (light grey) lines show the initial energies which are equal for all modes. Left-handed and right-handed modes are in the left and in the right columns, respectively.}
  \label{fig:E_power_spectra}
\end{figure*}

 On Figures~\ref{fig:B_power_spectra}~and~\ref{fig:E_power_spectra} we separately illustrate the time dependence of the energies of the left-handed (on the left) and right-handed (on the right) helical magnetic and electric fields with wave numbers $k \leq \frac{2 \pi n}{L_3}$ for several most characteristic sets of simulation parameters. The wave numbers are coded in color, from pure blue for the smallest nonzero value $k = \frac{2 \pi}{L_3}$ (largest wavelength) to pure red for $k = \frac{2 \pi n}{L_3}$. Semi-transparent colored regions show the range of short-scale oscillations of $I^{E,B}_{k,R/L}\lr{t}$, and thick solid lines show the time dependence of the time-smeared energies $\bar{I}_{k,R/L}\lr{t}$ defined in (\ref{smeared_energies_def}). Horizontal green lines show the initial energies which are equal for all helical components of electric and magnetic fields.

 From Fig.~\ref{fig:B_power_spectra} we see that in some simulations (parameter sets No.~{3 -- 7}) the energies of the helical components of magnetic field exhibit the expected signatures of the inverse cascade due to chiral plasma instability \cite{Boyarsky:12:1,Boyarsky:15:1,Kharzeev:15:1,Yamamoto:16:1}. Namely, the energy of a single longest-wavelength right-handed helical mode rapidly grows at early times and reaches some saturation limit at late evolution time, whereas the energies of all other modes decrease with time. As expected from the anomalous Maxwell equations (\ref{anomalous_maxwell}) with $\sigma_{CME}=\mu_A/\lr{2 \pi^2}$, increasing $\mu_A$ by a factor of two (to $\mu_A = 1.5$, parameter set No.~3, second row in Fig.~\ref{fig:B_power_spectra}) results in the growth of two right-handed modes. Comparing the data on Fig.~\ref{fig:B_power_spectra} and Fig.~\ref{fig:q5_vs_t}, we conclude that the growth of helical magnetic fields is not necessarily accompanied by the decay of the axial charge, and vice versa (see also Table \ref{tab:parameter_sets} for a summary of all simulations). Yet another observation which supports this conclusion is that in simulations on the smallest lattice, for which the axial charge exhibits most rapid decay, we have not found any signatures of the growing electromagnetic fields. Interestingly, increasing the value of $\mu_A$ and/or the number of initially excited electromagnetic field modes also does not necessarily speed up the inverse cascade and the decay of $Q_A$.

 An even more interesting picture emerges if we also consider the energies of the helical components of electric fields, shown on Fig.~\ref{fig:E_power_spectra}. It turns out that for some simulation parameters the long-wavelength helical components of the electric field, rather than the magnetic field, are enhanced during the evolution (parameter sets No.~{1, 2, 8}). For parameter set No.~3, both magnetic and electric fields grow in time. It is remarkable that precisely for these parameter sets the axial charge exhibits most rapid decay. It seems that both the growth of the helical electric fields and the decay of the axial charge are triggered by sufficiently large initial amplitudes of electromagnetic fields. Thus it seems that the roles of electric and magnetic fields in the chiral plasma instability scenario are essentially different, in contrast to the simple solution (\ref{exp_growing_solution}) of the anomalous Maxwell equations. It is also interesting to note that we observe the maximal growth of long-wavelength helical electric fields in simulations with a smaller value of Fermi velocity $v_F = 0.75$ (parameter set No.~8). Such a strong dependence on the Fermi velocity calls for a proper theoretical analysis.

 In order to quantify the net transfer of energy due to the inverse cascade, we follow \cite{Yamamoto:16:1} and introduce the magnetic and electric correlation lengths $\xi_{B}\lr{t}$ and $\xi_E\lr{t}$ as
\begin{eqnarray}
\label{elmag_corr_def}
 \xi_{E,B}\lr{t} = \frac{\sum\limits_{k} \frac{2 \pi}{k} I^{E,B}_k\lr{t} }{\sum\limits_k I^{E,B}_k\lr{t}},
\end{eqnarray}
where $I^{E,B}_k\lr{t} = I^{E,B}_{k,R}\lr{t} + I^{E,B}_{k,L}\lr{t}$. The time dependence of $\xi_E\lr{t}$ and $\xi_B\lr{t}$, shown on Fig.~\ref{fig:xEB_vs_t}, quantifies the direction of the transfer of energy between short- and long-wavelength modes. $\xi_B\lr{t}$ and $\xi_E\lr{t}$ can be also thought of as the average wavelengths of magnetic and electric fields at a given moment of time. The data shown on Fig.~\ref{fig:xEB_vs_t} indicates that $\xi_E$ and $\xi_B$ on average increase with time practically for all our simulations, thus providing a more quantitative evidence for the inverse cascade. The growth is somewhat more pronounced for the electric correlation length $\xi_E$, especially in simulations with larger initial amplitude $f = 0.2$. At late evolution times, $\xi_E$ saturates at its upper bound $\xi_E = L_3$ equal to the lattice size. In contrast, the magnetic correlation length $\xi_B$ exhibits rapid growth only at early times, and later seems to saturate at values smaller than $L_3$. On Fig.~\ref{fig:xEB_vs_t} we do not show the data for the parameter set No.~9, since in this case we have found that only a single initially excited mode strongly dominates the spectrum throughout the whole evolution process, and the quantities $\xi_E$ and $\xi_B$ are trivially equal to $L_3$ up to some very small corrections.

 The effect of saturation of $\xi_{E,B}$ at late evolution times prevents us from checking the universal late-time behavior $\xi_{E,B} \sim \sqrt{t}$ which follows from the scaling analysis of the anomalous Maxwell equations \cite{Kharzeev:15:1,Yamamoto:16:1}, similarly to (\ref{universal_scaling}). It seems, however, that the late-time behavior of $\xi_E$ is more similar to $\sqrt{t}$ than that of $\xi_B$.

 So far almost all theoretical studies of the chiral plasma instability assume that the axial charge is distributed homogeneously in space and can be described by a coordinate-independent chiral chemical potential $\mu_A$ at all evolution times. The extension of the anomalous Maxwell equations (\ref{anomalous_maxwell}) which allows to consider spatially inhomogeneous distributions of axial charge density has been constructed only recently in \cite{Shovkovy:16:1}. It is thus interesting to check how well the assumption of spatial homogeneity of the axial charge density $q_{A\,x}$ holds in our simulations. In order to quantify the spatial inhomogeneity of $q_{A\,x}$, we consider the space-averaged squared deviation of $q_{A\,x}$ from its space-averaged value $Q_A/V$:
\begin{eqnarray}
\label{q5dis_def}
 \sigma\lrs{q_A} = \sqrt{\sum\limits_x \lr{q_{A\,x} - Q_A/V}^2} .
\end{eqnarray}
The time dependence of the ratio $\sigma\lrs{q_A}/\lr{Q_A/V}$ is shown on Fig.~\ref{fig:q5dis_vs_t} for those sets of simulation parameters which exhibit axial charge decay (in particular, this fixes $f = 0.2$). Since the Hamiltonian which we use to define the initial state of our simulations involves the initial spatially inhomogeneous configuration $A_0$ of vector potential, even at the start of the evolution the axial charge density is slightly inhomogeneous, with deviations from mean value being of order of $5 \, \%$. As one can see from Fig.~\ref{fig:q5dis_vs_t}, at late evolution times the inhomogeneity of $q_{x\,A}$ tends to slightly increase, however, this increase is not dramatic and does not exceed $20 \%$. This suggests that the approximation of spatially homogeneous axial charge distribution is not unreasonable even when the long-wavelength modes are strongly enhanced and dominate the evolution. For simulations with $f = 0.05$ which do not exhibit the decay of the axial charge, the inhomogeneities of the axial charge density remain approximately constant or even tend to decrease.

 An interesting question is also the net transfer of energy between fermions and electromagnetic fields. As discussed in Section~\ref{sec:methods}, in classical statistical field theory algorithm the total energy of fermions and electromagnetic fields is conserved up to the work performed by the external current (see Fig.~\ref{fig:energy_joint} in Appendix~\ref{apdx:csft_derivation} for a numerical demonstration). Since in the simulations of chiral plasma instability discussed in this Section the external currents are absent, the transfer of energy can be characterized by the time dependence of the energy of electromagnetic field alone, which is illustrated on Fig.~\ref{fig:em_energy_vs_t}. We see that in almost all simulations the energy of electromagnetic field decreases or stays constant. The only exception is the simulation with parameter set No.~7 ($n = 10$, $f = 0.05$, $\mu_A = 1.5$), for which the energy of electromagnetic field quickly increases by almost a factor of three at $t \gtrsim 6 \cdot 10^3$. Analysis of power spectra suggests that this increase can be at least partly attributed to the enhancement of helical long-wavelength electric fields (similarly to the one observed for parameter set No.~3, see second row on Fig.~\ref{fig:E_power_spectra}). The decrease of electromagnetic field energy in all other simulations indicates that it might be not completely correct to think of chiral instability as of a ``discharge'' of on excited state of Dirac sea into electromagnetic waves.

\section{Conclusions}
\label{sec:conclusions}

 In this work, we have studied the real-time quantum evolution of chirally imbalanced Wilson-Dirac lattice fermions coupled to dynamical classical electromagnetic field within the classical statistical field theory approach. The quantum evolution of fermions was simulated exactly (up to small fully controlled errors originating from discretization of time). Our simulations of the chirality pumping process, described by the volume-integrated anomaly equation (\ref{volume_integrated_anomaly}), suggest that the effect of explicit chiral symmetry breaking due to the Wilson term in the lattice Dirac Hamiltonian is not very large. We hope therefore that our results can be confronted at least at the qualitative level with the theoretical predictions for continuum chiral fermions.

 We have considered both the generation of chirality imbalance in parallel electric and magnetic fields and the decay of initially present chirality imbalance at the expense of generating electromagnetic fields with nonzero helicity. We have observed that in general the backreaction of dynamical electromagnetic fields prevents fermions from acquiring large chirality imbalance - either by suppression of the chirality pumping or by accelerating the decay of initially present chirality imbalance. The suppression of the chirality pumping process can be understood as the dynamical screening of the external electric field, similarly to what happens in the Schwinger pair creation process \cite{Berges:14:1,Tanji:13:1}.

 In simulations with nonzero initial axial charge $Q_A$ we have also found a numerical evidence of the inverse cascade phenomenon due to the chiral plasma instability - that is, rapid growth of long-wavelength magnetic fields of definite helicity at early evolution times and the decay of all other magnetic field components. In some cases, helical electric fields were found to grow, even when magnetic fields did not exhibit any enhancement. A summary of our simulations given in Table \ref{tab:parameter_sets} suggests that the growth (or at least the absence of decay) of long-wavelength helical electric fields is a necessary condition for the dynamical decay of the axial charge. The fact that the enhancement of helical electric fields is switched on only for sufficiently large initial amplitude of electromagnetic field indicates that nonlinear responses such as the dynamical refringence \cite{Berges:16:2} might be important for the evolution of chirally imbalanced plasma.

 We have observed the mechanism which eventually stops the growth of long-wavelength modes in our simulations is not directly related to the decay of the axial charge. This observation, together with quite different roles of electric and magnetic fields in the evolution process, suggests that the nontrivial momentum and frequency dependence of both the electric conductivity and the chiral magnetic conductivity might be important for the quantitative description of chiral plasma instability. On the other hand, our simulations also indicate that the approximation of spatially homogeneous axial charge distribution, assumed in most theoretical considerations of anomalous Maxwell equations, is reasonably good even at late evolution times, when the instability has fully developed and the growth of long-wavelength helical electromagnetic fields has saturated.

 An interesting further development of our work would be to use chiral lattice fermions in the CSFT algorithm, with the possible choice of overlap Hamiltonian \cite{Creutz:01:1}. In this case, axial charge is conserved in the absence of electromagnetic fields, and the effects of explicit chiral symmetry breaking at high momenta should be absent. Such setup should be more relevant in the context of high-energy physics, where chiral symmetry tends to be exact at sufficiently high energies (at least at the level of the bare Lagrangian). Yet another interesting open question is the effect of the quantum fluctuations of the electromagnetic field, which are encoded in the nontrivial initial density matrix.

\begin{acknowledgments}
 We thank D.~Kharzeev, A.~Sadofyev and N.~Yamamoto for interesting and stimulating discussions of the physics of chiral media, and F.~Hebenstreit and D.~Gelfand for useful discussions of the CSFT algorithm. This work was supported by the S.~Kowalevskaja award from the Alexander von Humboldt Foundation. We are also indebted to S.~Valgushev for his help with code parallelization at the initial stage of this work as well as for the careful reading of this manuscript. The authors are grateful to FAIR-ITEP supercomputer center where a part of these numerical calculations was performed.
\end{acknowledgments}

\appendix

\section{Classical statistical field theory algorithm}
\label{apdx:csft_derivation}

 The starting point of our derivation of the CSFT algorithm is the general expression for the time-dependent expectation value of some quantum operator $\hat{O}$:
\begin{eqnarray}
\label{basic_ev}
 \vev{\hat{O}\lr{t}}
 =
 \tr\lr{ \hat{\rho}_0 \hat{U}\lr{t_0, t} \hat{O} \hat{U}^{\dag}\lr{t_0, t} } ,
\end{eqnarray}
where $\hat{\rho}_0$ is the initial density matrix and the evolution operator $\hat{U}\lr{t_0, t}$ is the time-ordered exponent
\begin{eqnarray}
\label{t_exp}
\hat{U}\lr{t_0, t} = \mathcal{T}\expa{ -i \int\limits_{t_0}^t dt' \hat{H}\lr{t'}  }  ,
\end{eqnarray}
where the Planck constant is set to one by an appropriate choice of units. The Hamiltonian operator is defined by equations (\ref{hferm_def}), (\ref{h_wd}) and (\ref{h_em}) in Section~\ref{sec:methods}. We have allowed for an explicit time dependence of the Hamiltonian, for example, due to the time dependence of the external current $\mathcal{J}_{x,i}\lr{t}$. The evolution operator (\ref{t_exp}) can be expanded into a product of elementary evolution operators for small time step $\delta=\frac{t-t_0}{N}$, where $\delta^{-1}$ should be much larger than any relevant energy scale in the system:
\begin{eqnarray}
\label{suzuki_dec}
 \hat{U}\lr{t, t_0}
 = \nonumber\\ =
 \lim_{\delta \rightarrow 0}
 \lr{ e^{-i \hat{H}(t_0) \delta}   e^{-i \hat{H}(t_0+\delta) \delta} \ldots   e^{-i \hat{H}(t) \delta}   }.
\end{eqnarray}

\begin{figure}[h!b]
  \centering
  \vspace{0.5cm}
  \includegraphics[width=6cm]{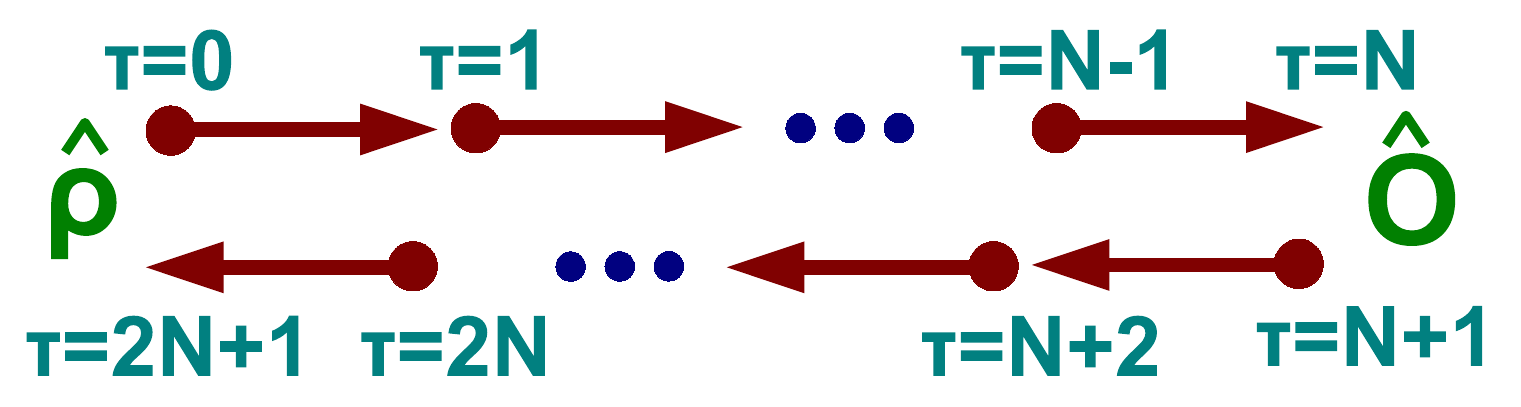}\\
  \caption{An illustration of the Schwinger-Keldysh contour with the discrete lattice time $\tau$.}
  \label{fig:schwinger_keldysh_contour}
\end{figure}

 Let us now insert the decompositions of the identity operator $\hat{I} = \prod\limits_{x,i} \int dA_{x, i} \ket{A_{x, i}} \bra{A_{x, i}}$ in the Hilbert space of electromagnetic field between the infinitesimal factors as well as at the beginning and at the end of the product in (\ref{suzuki_dec}) in order to arrive at the path integral representation of the evolution operator (\ref{t_exp}). We do this both for the forward and the backward evolution operators $\hat{U}\lr{t_0, t}$ and $\hat{U}^{\dag}\lr{t_0, t}$ in (\ref{basic_ev}). It is convenient to enumerate the gauge fields which enter identity decompositions in the forward evolution operators with the discrete lattice time variable $\tau = 0 \ldots N$, and in the backward branch - with $\tau = N+1 \ldots 2 N + 1$. The variable $\tau$ is a discrete parametrization of the Keldysh contour going from $t_0$ to $t$ and back (see Fig.~\ref{fig:schwinger_keldysh_contour} for an illustration). Now we have to express the matrix elements $\bra{A^{\tau}} e^{\mp i \hat{H}\lr{\tau} \delta} \ket{A^{\tau+1}}$ of the elementary evolution operators in terms of the fields $A^{\tau}$ and $A^{\tau+1}$. In the derivation of the CSFT algorithm, it is most convenient to use the approximate expression
\begin{eqnarray}
\label{evol_matr_el_forward}
 \bra{A^{\tau}} e^{- i \hat{H}\lr{\tau} \delta} \ket{A^{\tau+1}}
 \approx e^{+\frac{i}{2 \delta} \sum\limits_{x,i}\lr{A_{x,i}^{\tau+1} - A_{x,i}^{\tau}}^2}
 \times \nonumber \\ \times
 e^{
 - \frac{i \delta}{2} \sum\limits_{x,i,j} \lr{F_{x,i,j}^{\tau}}^2
 - i \delta \hat{H}_F\lrs{A^{\tau}}
 - i \delta \sum\limits_{x,i} A_{x,i}^{\tau} \mathcal{J}_{x,i}^{\tau}
 }
\end{eqnarray}
for the forward evolution operators, and the different approximate expression
\begin{eqnarray}
\label{evol_matr_el_backward}
 \bra{A^{\tau}} e^{+ i \hat{H}\lr{\tau} \delta} \ket{A^{\tau+1}}
 \approx e^{-\frac{i}{2 \delta} \sum\limits_{x,i}\lr{A_{x,i}^{\tau+1} - A_{x,i}^{\tau}}^2}
 \times \nonumber \\ \times
 e^{
 + \frac{i \delta}{2} \sum\limits_{x,i,j} \lr{F_{x,ij}^{\tau+1}}^2
 + i \delta \hat{H}_F\lrs{A^{\tau+1}}
 + i \delta \sum\limits_{x,i} A_{x,i}^{\tau+1} \mathcal{J}_{x,i}^{\tau+1}
 }
\end{eqnarray}
for the backward evolution operators. In the first expression (\ref{evol_matr_el_forward}), we order the electromagnetic field operators $\hat{E}_{x,i}$ and $\hat{A}_{x,i}$ in such a way that all the operators in the exponent containing $\hat{A}_{x,i}$ act on the vector $\bra{A^{\tau}}$. In the second expression (\ref{evol_matr_el_backward}), these operators act on the vector $\ket{A^{\tau+1}}$. These approximations are both valid to order $O(\delta)$ and differ only in the terms of order $O(\delta^2)$, hence being equivalent in the limit $\delta \rightarrow 0$.

 Using (\ref{evol_matr_el_forward}) and (\ref{evol_matr_el_backward}), we arrive at the path integral representation of the expectation value $\vev{\hat{O}\lr{t}}$, in which we integrate over the gauge fields living on the discretized Keldysh contour:
\begin{widetext}
\begin{eqnarray}
\label{Keldysh1}
\vev{ \hat{O}\lr{t} }
= \int dA^0 \ldots dA^{2N+1} \rho_{EM}\lrs{A^0, A^{2N+1}}
  \times \nonumber \\ \times
\Tr  \left( \hat{\rho}_F  \,
e^{+\frac{i}{2 \delta} \sum\limits_{x,i} \lr{A_{x,i}^1 - A_{x,i}^0}^2 }
e^{-i \delta \hat{H}_F\lrs{A^0}}
e^{-\frac{i\delta}{2} \sum\limits_{x,i,j} \lr{F_{x,ij}^0}^2 -i \delta \sum\limits_{x,i} A_{x,i}^0 \mathcal{J}_{x,i}^0 }
\right. \times \ldots \nonumber\\ \ldots \times
e^{+\frac{i}{2 \delta} \sum\limits_{x,i} \lr{A_{x,i}^N - A_{x,i}^{N-1}}^2 }
e^{-i \delta \hat{H}_F\lrs{A_{N-1}}}
e^{-\frac{i\delta}{2} \sum\limits_{x,i,j} \lr{F_{x,ij}^{N-1}}^2 - i \delta \sum\limits_{x,i} A_{x,i}^{N-1} \mathcal{J}_{x,i}^{N-1}}
\, \hat{O}\lrs{A^N, A^{N+1}} \,
\times \nonumber \\ \times
e^{-\frac{i}{2 \delta} \sum\limits_{x,i} \lr{A_{x,i}^{N+2} - A_{x,i}^{N+1}}^2 }
e^{i \delta \hat{H}_F\lrs{A^{N+2}}}
e^{\frac{i\delta}{2} \sum\limits_{x,i,j} \lr{F_{x,ij}^{N+2}}^2 +  i \delta \sum\limits_{x,i} A_{x,i}^{N+2} \mathcal{J}_{x,i}^{N-1}}
\times \ldots \nonumber\\ \left. \ldots \times
e^{-\frac{i}{2 \delta} \sum\limits_{x,i} \lr{A_{x,i}^{2N+1} - A_{x,i}^{2N}}^2 }
e^{i \delta \hat{H}_F\lrs{A^{2N+1}}}
e^{\frac{i\delta}{2} \sum\limits_{x,i,j} \lr{F_{x,ij}^{2N+1}}^2 + i \delta \sum\limits_{x,i} A_{x,i}^{2 N+1} \mathcal{J}_{x,i}^{0} }
\right)  .
\end{eqnarray}
\end{widetext}
It is important to stress that at this point we have used the path integral representation only for the bosonic fields, and the exponential factors $e^{\pm i \delta \hat{H}_F\lrs{A^{\tau}}}$ in (\ref{Keldysh1}) are still operators in the fermionic many-body Hilbert space. Correspondingly, the trace in (\ref{Keldysh1}) is taken over this Hilbert space. The operator of the observable $\hat{O}\lrs{A^N, A^{N+1}} = \bra{A^N} \hat{O} \ket{A^{N+1}}$ is also an operator on the fermionic Hilbert space which depends on fields $A^N$ and $A^{N+1}$. In deriving the above path integral representation, we have made a simplifying assumption \cite{Berges:14:1} that the initial density matrix $\hat{\rho}_0$ factorizes into the direct product of the fermionic density matrix $\hat{\rho}_F$ (which might in general be correlated with the initial state of the electromagnetic field) and the density matrix $\hat{\rho}_{EM}$ of the electromagnetic field with matrix elements $\rho_{EM}\lr{A^0, A^{2N+1}} = \bra{A^0} \hat{\rho}_{EM} \ket{A^{2N+1}}$. While this assumption is certainly not valid, say, for the density matrix $\hat{\rho} = e^{-\hat{H}/T}$ describing the thermal equilibrium state of the full Hamiltonian $\hat{H}_{EM} + \hat{H}_F$, it is still justifiable in the case of almost classical dynamics of electromagnetic fields.

 At this point let us assume that the observable operator $\hat{O}\lrs{A^N, A^{N+1}}$ can be represented as a sum of the identity operator in fermionic Hilbert space (this summand corresponds to purely bosonic observables) and of all possible fermionic bilinear operators:
\begin{eqnarray}
\label{fermionic_observable_def}
 \hat{O}\lr{A^N, A^{N+1}}
 = O_B\lr{A^N, A^{N+1}} \hat{I}
 + \nonumber \\ +
 \sum\limits_{x,y}
 \hat{\psi}^{\dag}_x
 \lrs{O_F\lr{A^N, A^{N+1}}}_{x,y}
 \hat{\psi}_y .
\end{eqnarray}
This form is sufficiently general to describe all the observables which we consider in this work. Furthermore, let us assume that the fermionic density matrix $\hat{\rho}_F$ can be represented as an exponent of some fermionic bilinear operator $\hat{H}_0 = \sum\limits_{x,y} \hat{\psi}^{\dag}_x \lrs{h_0}_{x,y} \hat{\psi}_y$:
\begin{eqnarray}
\label{fermionic_dm_bilinear}
 \hat{\rho}_F = \mathcal{Z}^{-1} \expa{-\hat{H}_0/T} ,
\end{eqnarray}
where $T$ is some (perhaps fictitious) temperature. Note that in the case of evolution which starts from non-equilibrium state the operator $\hat{H}_0$ can be different from the Wilson-Dirac Hamiltonian $\hat{H}_F$ which governs the quantum evolution. For instance, the excited state with initial chiral imbalance considered in Section~\ref{sec:chiral_instability} corresponds to the following form of $h_0$ in the limit $T \rightarrow 0$:
\begin{eqnarray}
\label{h0_noneq_filling}
 h_0 = h\lrs{A^0} + \mu_A \sum\limits_a \ket{\psi_a} \sign\lr{\bra{\psi_a} \gamma_5 \ket{\psi_a}}  \bra{\psi_a} ,
\end{eqnarray}
where $\ket{\psi_a}$ are the eigenstates of the Wilson-Dirac Hamiltonian (\ref{h_wd}) with the initial gauge field $A_0$. It is obvious that for exactly chiral Dirac Hamiltonian with $\bra{\psi_a} \gamma_5 \ket{\psi}_a = \pm 1$ this definition reduces to the form $h_0 = h + \mu_A \gamma_5$.

 Now we are in the position to further simplify the trace over the many-body fermionic Hilbert space in (\ref{Keldysh1}). To this end we use the identities
\begin{eqnarray}
\label{fermionic_trace_id}
 \tr\lr{ e^{\hat{B}_1} \ldots e^{\hat{B}_n} }
 =
 \det{1 + e^{B_1} \ldots e^{B_n}} ,
 \\ \label{fermionic_trace_bilinear}
 \tr\lr{ e^{\hat{B}_1} \ldots e^{\hat{B}_n} \hat{O}_F } =
 \det{1 + e^{B_1} \ldots e^{B_n}}
 \times \nonumber \\ \times
 \tr\lr{\lr{1 + e^{-B_n} \ldots e^{-B_1}}^{-1} O_F }  ,
\end{eqnarray}
where the operators $\hat{B}_i = \sum\limits_{x,y} \hat{\psi}^{\dag}_x \lrs{B_i}_{x,y} \hat{\psi}_y$ and $\hat{O}_F = \sum\limits_{x,y} \hat{\psi}^{\dag}_x O_{x,y} \hat{\psi}_y$ are the fermionic bilinear operators, and the corresponding symbols without hats denote operators on the single-particle fermionic Hilbert space with matrix elements $\lrs{B_i}_{x,y}$ and $O_{F \, x,y}$. Correspondingly, on the left hand side the traces are over the many-body fermionic Hilbert space, and the determinants and traces on the right hand side are on the single-particle fermionic Hilbert space.

 As yet another preliminary step in the derivation of the CSFT algorithm, let us also decompose the gauge fields on the forward and the backward branches of the Keldysh contour into the ``classical'' gauge field $\bar{A}_{x,i}^{\tau}$ and the ``quantum'' gauge field $\tilde{A}_{x,i}^{\tau}$ as
\begin{eqnarray}
\label{gauge_field_decomp}
 A_{x,i}^{\tau} = \bar{A}_{x,i}^{\tau} +  \frac{1}{2} \tilde{A}_{x,i}^{\tau},
 \nonumber \\
 A_{x,i}^{2N+1-\tau} = \bar{A}_{x,i}^{\tau} -  \frac{1}{2} \tilde{A}_{x,i}^{\tau},
 \quad \tau = 0 \ldots N .
\end{eqnarray}

 Relying on the assumptions (\ref{fermionic_observable_def}) and (\ref{fermionic_dm_bilinear}) and using the identities (\ref{fermionic_trace_id}) and (\ref{fermionic_trace_bilinear}), one can rewrite the expression (\ref{Keldysh1}) in terms of the operators on the single-particle fermionic Hilbert space and the variables $\bar{A}_{x,i}^{\tau}$ and $\tilde{A}_{x,i}^{\tau}$:
\begin{widetext}
\begin{eqnarray}
\label{Keldysh2}
\vev{ \hat{O}\lr{t} }
= \mathcal{Z}^{-1} \,
  \int d\bar{A}^0 \ldots d\bar{A}^{N}
  \int d\tilde{A}^0 ... d\tilde{A}^{N}
  \rho_{EM}\lr{\bar{A}^0 + \frac{\tilde{A}^0}{2}, \bar{A}^0 - \frac{\tilde{A}^0}{2}}
  \expa{
  \tr\ln\lr{
   1 + u_{-} e^{-h_0/T} u_{+}
  }}
  \times \nonumber \\ \times
  \expa{
  \frac{i}{\delta}
  \sum\limits_{\tau=0}^{N-1}
  \sum\limits_{x,i}
  \lr{\tilde{A}_{x,i}^{\tau+1} - \tilde{A}_{x,i}^{\tau}}
  \lr{\bar{A}_{x,i}^{\tau+1} - \bar{A}_{x,i}^{\tau}}
  - i \delta
  \sum\limits_{\tau=0}^{N-1}
  \sum\limits_{x, i}
  \tilde{A}_{x,i}^{\tau}
  \lr{\mathcal{J}_{x,i}^{\tau} + \sum\limits_j \bar{F}_{x,i,j}^{\tau} - \bar{F}_{x-\hat{j},i,j}^{\tau}}
  }
  \times \nonumber \\ \times
  \left(
  O_0\lr{\bar{A}^N + \frac{\tilde{A}^N}{2}, \bar{A}^N - \frac{\tilde{A}^N}{2}}
  +
  \tr\lr{\lr{1 + u_{+}^{-1} e^{+h_0/T} u_{-}^{-1}}^{-1}
  O_1\lr{\bar{A}^N + \frac{\tilde{A}^N}{2}, \bar{A}^N - \frac{\tilde{A}^N}{2}}
  }
  \right) ,
\end{eqnarray}
\end{widetext}
where the field strength tensor $\bar{F}_{x,ij}^{\tau}$ is constructed from the ``classical'' component of the gauge field $\bar{A}_{x,i}^{\tau}$ exactly in the same way as in (\ref{f_def}) and we have introduced the unitary single-particle forward and backward evolution operators
\begin{eqnarray}
\label{singlepart_evol_ops}
 u_{+}
 =
 e^{-i \delta h\lrs{\bar{A}^{0}     + \frac{\tilde{A}^{0}}{2}} }
   \ldots
   e^{-i \delta h\lrs{\bar{A}^{N-1} + \frac{\tilde{A}^{N-1}}{2}}}  ,
   \nonumber \\
 u_{-}
 = e^{+i \delta h\lrs{\bar{A}^{N-1} - \frac{\tilde{A}^{N-1}}{2}}}
   \ldots
   e^{+i \delta h\lrs{\bar{A}^{0}     - \frac{\tilde{A}^{0}}{2}}} .
\end{eqnarray}

 The path integral representation (\ref{Keldysh1}) is exact in the limit $N \rightarrow \infty$, $\delta \rightarrow 0$ with fixed $t = N \delta$ (up to the simplifying assumptions on the form of the observable operator $\hat{O}$ and the initial density matrix $\hat{\rho}$), but is not suitable for numerical analysis. The key step in the derivation of the CSFT algorithm is to expand the fermion-induced effective action of electromagnetic field $S_F = \tr\ln\lr{1 + u_{-} e^{-h_0/T} u_{+}}$ in the first line of (\ref{Keldysh2}) to the linear order in the ``quantum'' electromagnetic field $\tilde{A}_{x,i}$:
\begin{eqnarray}
\label{eff_action_expansion}
S_F \approx
 \left. S_F \right|_{\tilde{A}_{x,i}^{\tau} = 0}
 +
 \sum\limits_{\tau=0}^{N}\sum\limits_{x,i}
 \tilde{A}_{x,i}^{\tau}
 \left.
 \frac{\partial}{\partial  \tilde{A}_{x,i}^{\tau}}
 S_F
 \right|_{\tilde{A}_{x,i}^{\tau} = 0} .
\end{eqnarray}
In order to calculate the first derivative of $S_F$ over $\tilde{A}_{x,i}^{\tau}$, we use the identities
\begin{eqnarray}
\label{sp_evops_derivatives}
 \left. \frac{\partial}{\partial  \tilde{A}_{x,i}^{\tau}} \,
 u_{+}
 \right|_{\tilde{A}_{x,i}^{\tau} = 0}
 =
 -\frac{i \delta}{2}
 u\lr{0, \tau}
 j\lrs{\bar{A}^{\tau}}
 u\lr{\tau, N} ,
 \nonumber \\
 \left. \frac{\partial}{\partial  \tilde{A}_{x,i}^{\tau}} \,
 u_{-}
 \right|_{\tilde{A}_{x,i}^{\tau} = 0}
 =
 -\frac{i \delta}{2}
 u^{\dag}\lr{\tau, N}
 j\lrs{\bar{A}^{\tau}}
 u^{\dag}\lr{0, \tau} ,
\end{eqnarray}
where we have introduced the single-particle operator of the conserved electric current
\begin{eqnarray}
\label{current_spop_def}
 j_{x,i}\lrs{A} = \frac{\partial h\lrs{A}}{\partial A_{x, i}}
\end{eqnarray}
as well as the single-particle evolution operator in the background of the ``classical'' electromagnetic field $\bar{A}_{x,i}^{\tau}$:
\begin{eqnarray}
\label{class_unit_evol}
 u\lr{\tau_1, \tau_2}
 =
 e^{-i \delta h\lrs{\bar{A}^{\tau_1}}}
 \ldots
 e^{-i \delta h\lrs{\bar{A}^{\tau_2-1}}},
 \quad \tau_2 \ge \tau_1 .
\end{eqnarray}
The identities (\ref{sp_evops_derivatives}) are exact up to the order $O\lr{\delta^2}$, since in the derivatives of the forward and backward evolution operators we have used different orderings of the elementary evolution operator $e^{-i \delta h\lrs{\bar{A}^{\tau}}}$ and the current operator $j\lrs{\bar{A}^{\tau}}$.

 Using (\ref{sp_evops_derivatives}) and the relations $\left. u_+\lr{0, N} \right|_{\tilde{A}=0} = u\lr{0, N}$ and $\left. u_-\lr{0, N} \right|_{\tilde{A}=0} = u^{\dag}\lr{0, N}$, after some simple algebraic manipulations we can rewrite the derivative over $\tilde{A}_{x,i}^{\tau}$ in (\ref{eff_action_expansion}) as
\begin{eqnarray}
\label{eff_action_derivative}
\left.
 \frac{\partial}{\partial  \tilde{A}_{x,i}^{\tau}}
 S_F
 \right|_{\tilde{A}_{x,i}^{\tau} = 0}
 \equiv \vev{j_{x,i}^{\tau}}
 = \nonumber \\ =
 -i \delta
 \tr\left(
 \frac{1}{1 + e^{h_0/T}}
 %\times \right. \nonumber \\  \left. \times
 u\lr{0, \tau}
 j_{x,i}\lrs{\bar{A}^{\tau}}
 u^{\dag}\lr{0, \tau}
 \right)
\end{eqnarray}

 Now that our action (\ref{eff_action_expansion}) is assumed to be linear in the ``quantum'' field $\tilde{A}_{x,i}^{\tau}$ for ${\tau = 1 \ldots N-1}$, the quantum field $\tilde{A}_{x,i}^{\tau}$ can be integrated out in a straightforward way in the case of purely bosonic observables with $O_F \equiv 0$. The case of fermionic observables with nontrivial $O_F\lr{\bar{A}^N + \frac{\tilde{A}^N}{2}, \bar{A}^N - \frac{\tilde{A}^N}{2}}$ is more subtle, since in the path integral representation (\ref{Keldysh2}) the fermionic observable itself depends on $\tilde{A}_{x,i}^{\tau}$ (via the factor $\lr{1 + u_{+}^{-1} e^{+h_0/T} u_{-}^{-1}}^{-1}$ under the fermionic trace in the last line of (\ref{Keldysh2})). It is a common assumption in the derivation of the CSFT algorithm to neglect the $\tilde{A}_{x,i}^{\tau}$ dependence of the fermionic observables (see e.g. \cite{Berges:14:1}), which can be justified, e.g., if the relevant physical processes involve large number of virtual fermionic particles. In this case one can argue that the exponent of the effective action $S_F$ has a much stronger dependence on $\tilde{A}_{x,i}^{\tau}$ than the observable. A heuristic argument in favor of such assumption is that if one neglects the $\tilde{A}_{x,i}^{\tau}$ dependence of the fermionic observables, the expectation values of all fermionic bilinear operators take exactly the same form as the expectation values of the electric current (\ref{eff_action_derivative}) and the fermionic energy (see equation (\ref{energy_conservation}) below). Since these quantities are related to the observables characterizing the classical electromagnetic field via the inhomogeneous Maxwell equations and the energy conservation law, they are certainly also physical observables. While the $\tilde{A}^{\tau}$ dependence of the observable operator might still encode some interesting effects of the backreaction of measurements on the quantum evolution, taking it into account would presumably lead to a significant complication of the CSFT algorithm. For all these reasons, we also assume that the factor $\lr{1 + u_{+}^{-1} e^{+h_0/T} u_{-}^{-1}}^{-1}$ in (\ref{Keldysh2}) depends negligibly weakly on $\tilde{A}^{\tau}$ and replace it by $\lr{1 + u^{-1}\lr{0, N} e^{+h_0/T} u\lr{0, N}^{\dag -1}}^{-1}$.

 In order to integrate out the fields $\tilde{A}_{x,i}^0$ and $\tilde{A}_{x,i}^N$ at the endpoints of the Keldysh contour, it is convenient to introduce the Wigner transforms $\bar{\rho}_{EM}\lr{\bar{A}^0_{x,i}, \bar{E}^0_{x,i}}$ and $\bar{O}_{1,2}\lr{\bar{A}^N_{x,i}, \bar{E}^N_{x,i}}$ of the initial density matrix $\hat{\rho}_{EM}$ and the operators $O_{F,B}$ in (\ref{fermionic_observable_def}):
\begin{eqnarray}
\label{rho_em_wigner}
\rho_{EM}\lr{\bar{A}^0 + \frac{\tilde{A}^0}{2}, \bar{A}^0 - \frac{\tilde{A}^0}{2}}
 = \nonumber \\ =
\int d\bar{E}_{x,i}^0 \bar{\rho}_{EM}\lr{\bar{A}_{x,i}^{0}, \bar{E}_{x,i}^0} e^{i \sum_{x,i} \bar{E}_{x,i}^0 \tilde{A}_{x,i}^{0} }
\\
\label{obs_wigner}
O_{F,B}\lr{\bar{A}^N + \frac{\tilde{A}^N}{2}, \bar{A}^N - \frac{\tilde{A}^N}{2}}
 = \nonumber \\ =
\int d\bar{E}_{x,i}^N \bar{O}_{F,B}\lr{\bar{A}_{x,i}^{N}, \bar{E}_{x,i}^N} e^{ - i \sum_{x,i} \bar{E}_{x,i}^N \tilde{A}_{x,i}^{N} }  ,
\end{eqnarray}
where $\bar{E}_{x,i}^{\tau}$ is the ``classical'' electric field. We also note that the first sum over $\tau$ in the second line of (\ref{Keldysh2}) can be rewritten as
\begin{eqnarray}
\label{time_int_byparts}
 \sum\limits_{\tau=0}^{N-1}
  \sum\limits_{x,i}
  \lr{\tilde{A}_{x,i}^{\tau+1} - \tilde{A}_{x,i}^{\tau}}
  \lr{\bar{A}_{x,i}^{\tau+1} - \bar{A}_{x,i}^{\tau}}
  = \nonumber \\ =
  - \sum\limits_{\tau=1}^{N-1} \sum\limits_{x,i}\tilde{A}_{x,i}^{\tau}
  \lr{\bar{A}_{x,i}^{\tau+1} + \bar{A}_{x,i}^{\tau-1} - 2 \bar{A}_{x,i}^{\tau}}
  + \nonumber \\ +
 \sum\limits_{x,i} \lr{   \tilde{A}_{x,i}^N \lr{\bar{A}_{x,i}^N - \bar{A}_{x,i}^{N-1}}
  - \tilde{A}_{x,i}^0 \lr{\bar{A}_{x,i}^1 - \bar{A}_{x,i}^0} } .
\end{eqnarray}

 Finally, we are ready to integrate out the ``quantum'' electromagnetic field $\tilde{A}_{x,i}^{\tau}$, which leads to the following expression for the expectation value $\vev{\hat{O}\lr{t}}$:
\begin{widetext}
\begin{eqnarray}
\label{CSFT_eqfin}
 \vev{\hat{O}\lr{t}} =
 \int d\bar{E}^0 \, d\bar{E}^N
 \int d\bar{A}^0 \ldots d\bar{A}^{N}
 \tilde{\rho}_{EM}\lr{\bar{A}^0, \bar{E}^0}
 \times \nonumber \\ \times
 \delta\lrs{ \bar{E}^0 - \frac{\bar{A}^1 - \bar{A}^0}{\delta} - \delta \mathcal{R}^0 }
  \prod\limits_{\tau=1}^{N-1}
 \delta\lrs{\frac{\bar{A}^{\tau+1} + \bar{A}^{\tau-1} - 2 \bar{A}^{\tau}}{\delta} + \delta \mathcal{R}^{\tau} }
 \delta\lrs{ \bar{E}^N - \frac{\bar{A}^N - \bar{A}^{N-1}}{\delta}}
 \times \nonumber \\ \times
 \lr{
 \bar{O}_{0}\lr{\bar{A}^N, \bar{E}^N}
 +
 \tr\lr{\lr{1 + e^{h_0/T}}^{-1} u\lr{0, N}
  \bar{O}_1\lr{\bar{A}^N, \bar{E}^N}
 u^{\dag}\lr{0, N}
 }
 },
\end{eqnarray}
\end{widetext}
where
\begin{eqnarray}
\label{phi_def}
\mathcal{R}_{x,i}^{\tau} =
\mathcal{J}_{x,i}^{\tau}
+
\langle \hat j_{x,i}^{\tau} \rangle
+
\sum_j \lr{\bar{F}_{x,ij}^{\tau} - \bar{F}_{x-\hat{j},ij}^{\tau}},
\end{eqnarray}
and $\vev{j_{x,i}^{\tau}}$ is the expectation value of the electric current defined as in (\ref{eff_action_derivative}). We note that the normalization factor $\mathcal{Z}^{-1}$ in (\ref{fermionic_dm_bilinear}) and (\ref{Keldysh2}) is cancelled by the zeroth-order term of the expansion (\ref{eff_action_expansion}).

 From the explicit expression (\ref{eff_action_derivative}) for the fermionic electric current $\vev{j_{x,i}^{\tau}}$  one can immediately see that it depends only on the classical electromagnetic field $\bar{A}_{x,i}^{\tau'}$ with $\tau' < \tau$. Therefore the delta-functions in the integral (\ref{CSFT_eqfin}) can be regarded as the constraints on the deterministic evolution of the classical electromagnetic field $\bar{A}_{x,i}^{\tau}$ interacting with the quantum fermionic field. To make this more obvious, we can rewrite the chain of $\delta$-functions in (\ref{CSFT_eqfin}) as
\begin{eqnarray}
\label{delta_func_chain}
 \delta\lrs{\bar{A}^1 - \mathcal{A}^1\lrs{\bar{A}^0, \bar{E}^0} }
 \times \nonumber \\ \times
 \prod\limits_{\tau=2}^{N} \delta\lrs{\bar{A}^{\tau} - \mathcal{A}^{\tau}\lrs{\bar{A}^{\tau-1}, \bar{A}^{\tau-2}, \vev{j^{\tau-1}} }  }
 \times \nonumber \\ \times
 \delta\lrs{\bar{E}^N - \frac{\bar{A}^N - \bar{A}^{N-1}}{\delta}},
 \nonumber \\
 \mathcal{A}^{1}\lrs{\bar{A}^0, \bar{E}^0} = \bar{A}^0 + \delta \, \lr{\bar{E}^0 - \delta\mathcal{R}^0} ,
 \nonumber \\
 \mathcal{A}^{\tau}\lrs{\bar{A}^{\tau-1}, \bar{A}^{\tau-2}, \vev{j^{\tau-1}}}
 = \nonumber \\ =
 2 \bar{A}^{\tau-1} - \bar{A}^{\tau-2} - \delta^2 \mathcal{R}^{\tau-1} ,
\end{eqnarray}
where the last definition is for ${\tau = 2 \ldots N}$. From this expression one can see that one can sequentially integrate out the fields $\bar{A}^{\tau}$ with ${\tau = 1 \ldots N-1}$ and express the fields $\bar{A}^N$, $\bar{E}^N$ in terms of the initial values $\bar{A}^0$, $\bar{E}^0$. Namely, integrating out the field $\bar{A}^1$ first, we remove the first delta-function in the product in (\ref{delta_func_chain}) and replace $\bar{A}^1$ by $\mathcal{A}^1\lrs{\bar{A}^0, \bar{E}^0}$ in the arguments of all the other delta functions. Integrating out $\bar{A}^2$, we remove the second delta-functions and replace $\bar{A}^2$ by ${\mathcal{A}^{2}\lrs{\bar{A}^0, \bar{E}^0} \equiv \mathcal{A}^{2}\lr{\mathcal{A}^1\lrs{\bar{A}^0, \bar{E}^0}, \bar{A}^0, \vev{j^1} }}$. We can repeat this process for all $\tau$ up to $N-1$, each time expressing $\mathcal{A}^{\tau}$ in terms of the initial values $\bar{A}^0$ and $\bar{E}^0$ and the functionals $\mathcal{A}^{\tau'}$ with $\tau' < \tau$. It is important that in such a sequential integration, the integrand $\bar{A}^{\tau}$ always enters the argument of the delta-function being removed linearly. Therefore despite the nonlinearity of the chain of evolution equations, such intermediate integrations do not produce any nontrivial Jacobian. To our knowledge, the absence of Jacobian in the integration measure in the CSFT algorithm so far has only been demonstrated for scalar field theory \cite{Jeon:05:1,Gozzi:99:1}. It is nice to see here explicitly its absence for lattice gauge theory coupled to fermions.

 After integrating out all the intermediate fields $\bar{A}^{\tau}$ with ${\tau = 1 \ldots N-1}$, we are left with the following form of equation (\ref{CSFT_eqfin})
\begin{eqnarray}
\label{CSFT_eqfin1}
 \vev{\hat{O}\lr{t}} =
 \int d\bar{A}^0 \, d\bar{E}^0
 \int d\bar{A}^N \, d\bar{E}^{N}
 \bar{\rho}_{EM}\lr{\bar{A}^0, \bar{E}^0}
 \times \nonumber \\ \times
 \delta\lrs{\bar{A}^N - \mathcal{A}^N\lrs{\bar{A}^0, \bar{E}^0} }
 \delta\lrs{\bar{E}^N - \mathcal{E}^N\lrs{\bar{A}^0, \bar{E}^0} }
 \times \nonumber \\ \times
 \left(
 \bar{O}_{0}\lr{\bar{A}^N, \bar{E}^N}
 + \right. \nonumber \\ \left. +
 \tr\lr{\frac{1}{1 + e^{h_0/T}} u\lr{0, N}
  \bar{O}_1\lr{\bar{A}^N, \bar{E}^N}
 u^{\dag}\lr{0, N}
 }
 \right) ,
\end{eqnarray}
where
\begin{eqnarray}
\label{delta_func_remainder}
 \mathcal{E}^N\lrs{\bar{A}^0, \bar{E}^0} =
 \frac{\mathcal{A}^N\lrs{\bar{A}^0, \bar{E}^0} - \mathcal{A}^{N-1}\lrs{\bar{A}^0, \bar{E}^0}}{\delta}
\end{eqnarray}
and we have expressed the functionals $\mathcal{A}^N$ and $\mathcal{A}^{N-1}$ in terms of the initial values $\bar{A}^0$, $\bar{E}^0$ of the vector gauge potential and the electric field. In this expression, it is straightforward to integrate out $\bar{A}^N$ and $\bar{E}^N$, which amounts to substituting $\mathcal{A}^N\lrs{\bar{A}^0, \bar{E}^0}$ and $\mathcal{E}^N\lrs{\bar{A}^0, \bar{E}^0}$ in place of $\bar{A}^N$ and $\bar{E}^N$ in the Wigner transforms of the observable operators $O_B$ and $O_F$.

 To summarize, the CSFT algorithm amounts to a simultaneous time evolution of the classical electromagnetic fields, described by the vector potential $\bar{A}_{x,i}^{\tau}$ and the electric field $\bar{E}_{x,i}^{\tau}$, and the quantum fermionic fields, described by single-particle evolution operator $u\lr{0, \tau}$ in (\ref{class_unit_evol}). The discrete equations which govern this evolutions (arguments of the delta functions in (\ref{CSFT_eqfin})) have a well-defined continuum limit $\delta \rightarrow 0$, $N \rightarrow \infty$ with fixed $t = N \delta$, which is given by the equations (\ref{continuum_maxwell}), (\ref{fermionic_current_vev}) and (\ref{single_particle_schrodinger}) in the main text. To simplify the notation, in the main part of the text we omit the bar over the classical components of the gauge field and denote them as $A_{x,i}\lr{t} \equiv \bar{A}_{x,i}^{\tau}$, $E_{x,i}\lr{t} \equiv \bar{E}_{x,i}^{\tau}$, $F_{x,ij}\lr{t} \equiv \bar{F}_{x,ij}^{\tau}$.

 In practice, however, the numerical solution of the continuum equations (\ref{continuum_maxwell}), (\ref{fermionic_current_vev}) and (\ref{single_particle_schrodinger}) should necessarily involve some discretization of time. While the simple discretization of the Keldysh contour used in the above derivation can be in principle used for numerical solution at sufficiently small $\delta$, for a given finite $\delta$ one can construct different, more advanced discretizations which would reduce discretization errors, thus improving the conservation of energy (\ref{energy_conservation}) and making the single-particle evolution operator $u\lr{0, \tau}$ numerically closer to a unitary matrix.

 In this work we follow \cite{Borsanyi:09:1,Berges:14:1} and use the leapfrog evolution scheme for the single-particle evolution operator $u^{\tau} \equiv u\lr{0, \tau}$, which significantly improves the conservation of the unitarity condition $u\lr{0, \tau} u^{\dag}\lr{0, \tau} = 1$ at finite discrete time step $\delta$:
\begin{eqnarray}
\label{modes_leapfrog}
 u^{\tau+1} = u^{\tau-1} - i \delta h\lrs{\bar{A}^{\tau}} u^{\tau}, \quad \tau = 1 \ldots N - 1
 \nonumber \\
 u^{1}      = u^0 - i \delta  h\lrs{\bar{A}^0} u^0, \quad u^0 = 1 .
\end{eqnarray}
In practice it is convenient to work with the components of $u^{\tau}$ in the basis of eigenstates of the initial single-particle Hamiltonian $h\lrs{\bar{A}^0}$. In particular, if translational invariance along some of the space directions is preserved during the evolution, $u^{\tau}$ remains block-diagonal in the basis of plane waves propagating along these directions. This block-diagonal structure can be used to greatly reduce the number of independent components of $u^{\tau}$ which enter the equations (\ref{modes_leapfrog}). We have used translational invariance in two out of three spatial lattice directions to speed up the evolution algorithm on large lattices with sizes up to $200 \times 40 \times 40$, assuming translational invariance in two out of three spatial directions.

 For the evolution of electromagnetic field we use the equations which directly follow from (\ref{CSFT_eqfin}):
\begin{eqnarray}
\label{em_fields_leapfrog}
 \frac{\bar{E}_{x,i}^{\tau+1} - \bar{E}_{x,i}^{\tau}}{\delta}
 = \nonumber \\ =
 -
 \mathcal{J}_{x,i}^{\tau}
 -
\langle \hat{j}_{x,i}^{\tau} \rangle
-
\sum\limits_j \lr{\bar{F}_{x,ij}^{\tau} - \bar{F}_{x-\hat{j},ij}^{\tau}} ,
\nonumber \\
\frac{\bar{A}_{x,i}^{\tau+1} - \bar{A}_{x,i}^{\tau}}{\delta} = \bar{E}_{x,i}^{\tau+1} ,
\nonumber \\
 \frac{\bar{A}^1_{x,i} - \bar{A}_{x,i}^0}{\delta} = \bar{E}^0_{x,i}
 - \nonumber \\ -
 \delta \lr{
 \mathcal{J}_{x,i}^{0}
 +
\langle \hat{j}_{x,i}^{0} \rangle
+
\sum\limits_j \lr{\bar{F}_{x,ij}^{0} - \bar{F}_{x-\hat{j},ij}^{0}}} .
\end{eqnarray}
In our simulations, we use the value $\delta = 0.05$. We have checked that decreasing $\delta$ down to $0.02$ does not change our results up to some small unimportant fluctuations.

 In principle, leapfrog-type time discretization (\ref{modes_leapfrog}) allows the existence of fermionic doublers in time direction - that is, the symmetric finite differences in (\ref{modes_leapfrog}) are zero if the mode functions oscillate as $\lr{-1}^{\tau}$. Such doubler modes correspond to another flavour of Dirac fermions with an opposite signature of the $\gamma_5$ matrix. Thus if such modes are excited, they can also contribute to the anomaly equation (\ref{volume_integrated_anomaly}) and effectively decrease the anomaly coefficient, or lead to the decay of the initial value of the axial charge \cite{Aarts:98:1}. In order to check whether fermionic modes with such high frequencies are excited we have calculated the average norm of forward finite differences of $u^{\tau}$ as $\frac{1}{4 V} \tr\lr{\lr{u^{\tau+1}-u^{\tau}}^{\dag} \lr{u^{\tau+1}-u^{\tau}}}$. Since the size of the single-particle Hilbert space is equal to $4 V$ this quantity should be of order of $\delta^2$ if $u^{\tau}$ are smooth functions of $\tau$. On the other hand, doubler modes with $u^{\tau} \sim (-1)^\tau$ yield the contribution of order of unity. In our simulations we have checked that the above norm remains of order of $10^{-2}$ for all evolution times and does not exhibit any tendency to grow. This suggests that the doubler modes remain practically unexcited during the evolution.

\begin{figure}
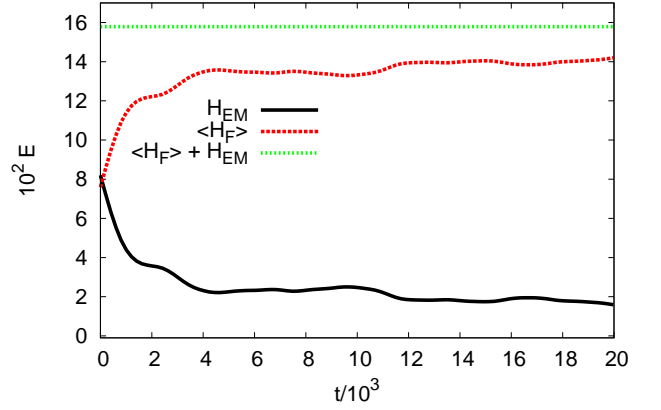

  \centering
  \includegraphics[width=6cm,angle=-90]{{{energy_joint}}}\\
  \caption{Time dependence of the energies of fermions and electromagnetic fields and their total for simulations on $200 \times 20 \times 20$ lattice with $n = 10$, $f = 0.2$, $\mu_A = 0.75$, $v_F = 1$ (parameter set No.~1 in Table \ref{tab:parameter_sets}).}
  \label{fig:energy_joint}
\end{figure}

 Another important characteristic of the discretization of the evolution equations (\ref{continuum_maxwell}) and (\ref{single_particle_schrodinger}) is the precision with which the conservation of energy (\ref{energy_conservation}) holds. For the leap-frog equations (\ref{modes_leapfrog}) and (\ref{em_fields_leapfrog}) the total energy of electromagnetic fields and fermions is conserved only approximately, up to the terms of order of $\delta^2$. In order to illustrate the conservation of energy in our simulations, on Fig.~\ref{fig:energy_joint} we show the time dependence of the fermionic energy $\vev{\hat{H}_F}$, the energy $H_{EM}$ of electromagnetic fields and their total. One can see that while both $\vev{\hat{H}_F}$ and $H_{EM}$ change quite strongly during the evolution, their sum is conserved with a very good precision, which again shows that the time step $\delta = 0.05$ is small enough.

 The discrete evolution equations (\ref{modes_leapfrog}) and (\ref{em_fields_leapfrog}) are ideally suited for parallelization on multi-node computers. Indeed, the largest amount of computer time is required to solve the evolution equation (\ref{modes_leapfrog}) for the $\lr{4 V} \times \lr{4 V}$ matrix $u^{\tau}$. Taking into account that only half of the single-particle fermionic states are filled at zero temperature, this size can be reduced by a factor of two down to $\lr{2 V} \times \lr{4 V}$. The evolution of electromagnetic field (\ref{em_fields_leapfrog}) requires only the total electric current summed over all fermionic modes, and is computationally very cheap. Thus it is natural to distribute the rows of the $u^{\tau}$ matrix over multiple nodes. On each node, one performs the elementary evolution step (\ref{modes_leapfrog}) for the rows attributed to this node and calculates the partial traces of the electric current and other fermionic bilinear operators over these rows. The results are sent to the master node, which calculates the total current and performs the evolution of the electromagnetic field according to (\ref{em_fields_leapfrog}). Since the amount of data transferred by network from each slave node to the master node is significantly smaller than the amount of data stored at each slave mode (for realistic lattice sizes and node numbers, the number of rows of $u^{\tau}$ per node is large), the speed of the algorithm scales practically linearly with the number of slave nodes.

 Such parallelization also solves the problem with very large amount of RAM memory required to store the matrix $u^{\tau}$ ($\sim 36$ Gb for the $20 \times 20 \times 20$ lattice when one uses 8-byte double accuracy numbers for all fields), which is simply split over different nodes. In order to further decrease the required RAM size, we use the 4-byte float numbers to store $u^{\tau}$. We have explicitly checked that the reduction from double to float real numbers practically does not affect our results.

 The extensive parallelization also allows us to avoid the stochastic summation over all modes \cite{Borsanyi:09:1,Tanji:13:1,Berges:13:1,Berges:14:1}, which introduces additional statistical noise in the results and can therefore significantly affect potentially unstable evolution which we study in this work. Instead, we perform exact summation over all initially occupied fermionic states.

%\bibliography{Buividovich}
%\bibliographystyle{apsrev}

\end{document}